\documentclass[twocolumn,aps,superscriptaddress,prb]{revtex4-2}

\usepackage{bm}%
\usepackage{float}
\expandafter\ifx\csname package@font\endcsname\relax\else
 \expandafter\expandafter
 \expandafter\usepackage
 \expandafter\expandafter
 \expandafter{\csname package@font\endcsname}%
\fi

\usepackage{comment}
\usepackage{array}
\usepackage{amsmath,amssymb}
\usepackage{MnSymbol}
\usepackage{tikz-feynman,contour} 
\usetikzlibrary{shapes,arrows,positioning,automata,backgrounds,calc,er,patterns}
\usepackage{feynmf}
\tikzfeynmanset{compat=1.0.0} 
 
\usepackage{graphicx}
\usepackage{mathrsfs}
\usepackage{bm}
\usepackage{mathtools}
\usepackage{epstopdf}
\usepackage{hyperref}

\hypersetup{%
   pdfpagemode=None, %FullScreen,
   pdfstartpage=1,
   pdfmenubar=true,
   pdftoolbar=true,
   colorlinks = true,
   linkcolor=blue,
   citecolor=blue,
   urlcolor=blue,
   bookmarksopen=false
 }
 \usepackage{amsmath}
\usepackage{amsfonts}
\usepackage{mathtools}
\usepackage{graphicx}
\usepackage{fixme}
\usepackage{color}
\usepackage{cancel}
\usepackage{bm}% bold math

\usepackage{epstopdf}
\usepackage{float}
\usepackage{amsmath}
\usepackage{color}
\usepackage{comment}
\usepackage{dsfont}
\usepackage{braket}
\usepackage[normalem]{ulem}

\newcommand\identity{1\kern-0.25em\text{l}}

\usepackage{feynmf}
\usepackage{tikz-feynman}
\usetikzlibrary{shapes,arrows,positioning,automata,backgrounds,calc,er,patterns}
\tikzfeynmanset{compat=1.0.0}

\newcommand{\Up}{{\uparrow}}
\newcommand{\Dn}{{\downarrow}}

\newcommand\bwt         {\begin{widetext}}
\newcommand\ewt         {\end{widetext}}

\def\ba{{\bf a}}
\def\bi{{\bf i}}
\def\bj{{\bf j}}
\def\rd{{\rm d}}
\def\bk{{\bf k}}
\def\bp{{\bf p}}
\def\bq{{\bf q}}
\def\bK{{\bf K}}
\def\bQ{{\bf Q}}

\def\ro{{\rm o}}
\def\rd{{\rm d}}
\def\br{{\bf r}}
\def\bd{{\bf d}}

\def\AF{{\rm AF}}
\def\mV{\mathcal{V}}
\newcommand{\bracket}[2]{\langle#1|#2\rangle}

\definecolor{airforceblue}{rgb}{0.36, 0.54, 0.66}
\definecolor{amber}{rgb}{1.0, 0.75, 0.0}
\definecolor{applegreen}{rgb}{0.55, 0.71, 0.0}
\definecolor{alizarin}{rgb}{0.82, 0.1, 0.26}

\begin{document}

\title{Wave function and spatial structure of   polarons in an antiferromagnetic  bilayer}

\author{J. H.\ Nyhegn}
\affiliation{Center for Complex Quantum Systems, Department of Physics and Astronomy, Aarhus University, Ny Munkegade, DK-8000 Aarhus C, Denmark}
\author{G. M.\ Bruun}
%\email[]{bruungmb@phys.au.dk}
\affiliation{Center for Complex Quantum Systems, Department of Physics and Astronomy, Aarhus University, Ny Munkegade, DK-8000 Aarhus C, Denmark}
%\affiliation{Shenzhen Institute for Quantum Science and Engineering and Department of Physics, Southern University of Science and Technology, Shenzhen 518055, China}
\author{K. Knakkergaard Nielsen}
\affiliation{Max-Planck Institute for Quantum Optics, Hans-Kopfermann-Str. 1, D-85748 Garching, Germany}

\begin{abstract}
Adding a dopant to an antiferromagnetic spin background disturbs the magnetic order and leads to the formation of a quasiparticle coined the magnetic polaron, which plays a central role in understanding strongly correlated materials. Recently, remarkably detailed insights into the spatial properties of such polarons have been obtained using atoms in optical lattices. Motivated by this we develop a nonperturbative scheme for calculating the wave function of the magnetic polaron in a bilayer antiferromagnet using the self-consistent Born approximation. The scheme includes an infinite number of spin waves, which is crucial for an accurate description of the most interesting regime of strong correlations. Utilizing the developed wave function, we explore the spatial structure of the polaron dressing cloud consisting of magnetically frustrated spins surrounding the hole. Mimicking the nonmonotonic behavior of the antiferromagnetic order, we find that the dressing cloud first decreases and then increases in size with increasing interlayer hopping. The increase reflects the decrease in the magnetic order as a quantum phase transition to a disordered state is approached for large interlayer hopping. We, furthermore, find that the symmetry of the ground state dressing cloud changes as the interlayer coupling increases. Our results should be 
experimentally accessible using quantum simulation with optical lattices.
\end{abstract}

\maketitle

\section{Introduction} 
The properties of dopants in antiferromagnetic (AFM) environments have been intensely studied both experimentally and theoretically for decades. The reasons for this are at least two-fold. First, the dopants are believed to be the charge carriers and fundamental building blocks for understanding 
 unconventional superconducting phases in cuprates~\cite{kane1989}. Second, the superconducting phases in cuprates as well as in pnictides \cite{wen2011}, organic layers \cite{wosnitza2012}, and twisted bilayer graphene \cite{cao2018} all reside close to the antiferromagnetically ordered phase. The mobile charge carriers can, therefore, provide important clues to the underlying physics behind unconventional superconducting phases. 
 The essential properties of these systems are captured by the Fermi-Hubbard model and the associated $t$--$J$ model. 
 Breakthroughs using quantum simulation of the Fermi-Hubbard model with ultracold atoms~\cite{esslinger2010,boll2016,cheuk2016b,mazurenko2017,hilker2017,brown2017,chiu2018,brown2019,chiu2019,brown2019,koepsell2019a,ji2021,koepsell2021} have stimulated new interest in this fundamental topic~\cite{Carlstrom2016,nagy2017,grusdt2018,grusdt2018a,nielsen2021,nielsen2022}. In particular, the
 impressive single-site resolution \cite{bakr2009,sherson2010,haller2015,yang2021} of such experiments makes it possible to measure spatial correlations, giving insights into the spatial structure of the emergent quasiparticles~
  \cite{koepsell2019a} and their dynamical formation~\cite{ji2021}, which are well beyond the reach of condensed matter experiments. 
  Recently, progress towards studying bilayer systems using optical lattices has been reported~\cite{gall2021}. This is not only interesting due to the relation to cuprates with superconducting blocks of two CuO$_{2}$ planes, but also because bilayer systems exhibit a quantum phase transition to a disordered state of interlayer spin singlets \cite{hida1992,sandvik1994,scalettar1994,millis1994a,sandvik1995,chubukov1995}. 

%%%%%%%%%%%%%%%%%%%%%%%%%%%%%%%%%%%%%%%%%%%%%%%
\begin{figure}[t!]
	\begin{center}
	\includegraphics[width=0.4\textwidth]{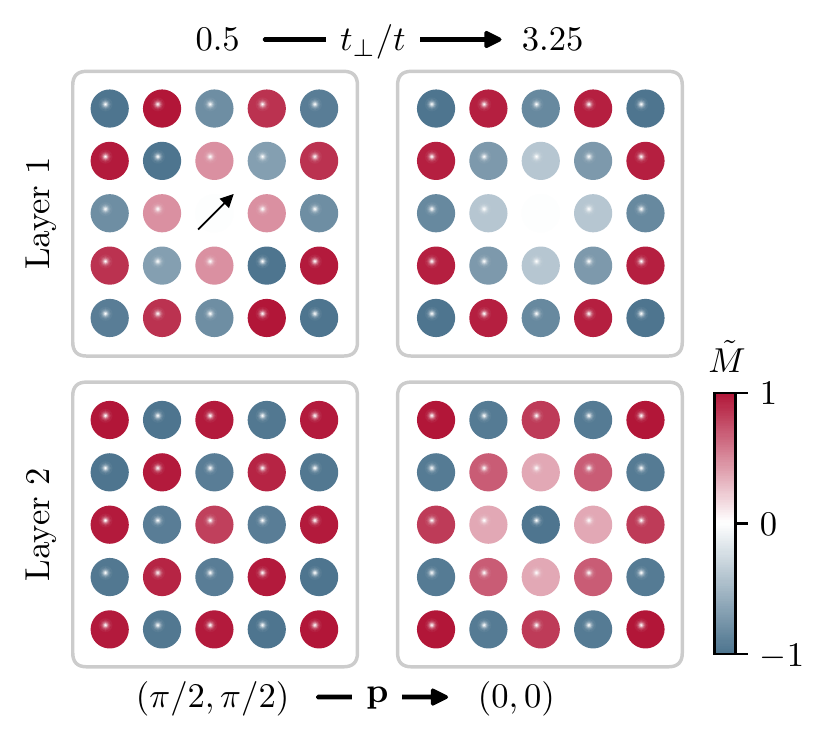}
	\end{center}
	\caption{ The magnetic dressing cloud of the polaron ground state for $t_{\perp}/t=0.5$ (left) and $t_{\perp}/t=3.25$ (right) with a hole in the center of layer 1. Here, $\tilde{M}=\pm1$ corresponds to the average magnetization predicted by linear spin wave theory (LSWT). For $t_{\perp}/t=0.5$, the ground state momentum is $\bp =(\pi/2,\pi/2)$ indicated by the arrow, and the cloud has $C_{2v}$ symmetry. Increasing $t_{\perp}/t$ to $3.25$ changes the ground state momentum to $\bp ={\bf 0}$, where the magnetization cloud recovers the full $C_{4v}$ symmetry of the antiferromagnet. In this limit, the spins in the dressing clouds in the two layers are approximately antiparallel. }
	\label{Fig:Front}
\end{figure}
%%%%%%%%%%%%%%%%%%%%%%%%%%%%%%%%%%%%%%%%%%%%%%%

On the theory side, the self-consistent Born approximation (SCBA) has proven to yield accurate descriptions of single dopants in these systems, whether it be their equilibrium \cite{martinez1991,liu1991,marsiglio1991,ramsak1998,chernyshev1999,wohlfeld2008,diamantis2021,nielsen2021} or nonequilibrium properties \cite{nielsen2022}. The ability to describe the spatial structure of the emergent magnetic polarons \cite{ramsak1998,nielsen2021} hinges in this framework on the analytical expression for the many-body wave function~\cite{reiter1994b,nielsen2022}. 

Motivated by the exciting progress regarding the experimental study of bilayer systems, we generalize the SCBA approach for a single layer \cite{reiter1994b,ramsak1998, nielsen2021}, and derive an analytical expression for the magnetic polaron many-body wave function for the bilayer geometry.
The wave function includes an infinite number of spin waves, which is important in the strongly correlated regime. 
This, in turn, allows us to calculate the magnetic dressing cloud around the hole by self-consistently solving a set of Dyson-like equations. We use this to explore how the dressing cloud is modified by the interlayer interaction. We find that the size of the ground state dressing cloud exhibits a nonmonotonic behavior with increasing hopping between the two layers and that it, additionally, changes its spatial symmetry (see Fig. \ref{Fig:Front}). 

The paper is organized as follows. In Sec. \ref{sec:model}, we introduce the $t$-$J$ model and discuss how we retrieve an effective model in terms of the hole and spin waves. The SCBA is then described and the quasiparticle wave function is constructed in Sec. \ref{sec:QP}. Using the quasiparticle wave function, we derive a set of Dyson-like equations used to calculate the spin frustration surrounding the hole in Sec. \ref{sec.spatial_structure}. In Sec. \ref{sec.spatial_structure_inter-layer_influence}, we explore the spatial structure of the quasiparticle as the interlayer coupling is varied before we conclude in Sec.~\ref{conclusions}.

\section{Bilayer $t$--$J$ model} \label{sec:model}
To model the behavior of a single hole in an AFM bilayer, we use the $t$--$J$ model with the Hamiltonian 
\begin{equation}
\hat{H}=\hat{H}_t+\hat{H}_J ,
\label{Eq:TotalH}
\end{equation} 
where %~\cite{brinkman1970}\georg{Bedre ref.?}
\begin{equation}
\hat{H}_t=-t\sum_{l,\braket{{\bf i}, {\bf j}},\sigma} \tilde{c}^{\dagger}_{l,{\bf i},\sigma} \tilde{c}_{l,{\bf j},\sigma}    
-t_\perp\sum_{{\bf i},\sigma} \tilde{c}^{\dagger}_{1,{\bf i},\sigma} \tilde{c}_{2,{\bf i},\sigma} +\text{H.c.} 
\label{Eq:Ht}
\end{equation}
is the hopping Hamiltonian, and 
\begin{align}
\hat{H}_J = & \ J\sum_{l,\braket{{\bf i}, {\bf j}}} \left[ \hat{\mathbf S }_{l,{\bf i}}\cdot \hat{\mathbf S }_{l,{\bf j}} -\frac{1}{4}\hat{n}_{l,\bi}n_{l,\bj} \right]  \nonumber \\ 
& \ +J_{\perp}\sum_{{\bf i}}  \left[  \hat{\mathbf S }_{1,{\bf i}}\cdot \hat{\mathbf S }_{2,{\bf i}} -\frac{1}{4}\hat{n}_{1,\bi}\hat{n}_{2,\bi} \right]
\label{Eq:HJ}
\end{align}
describes spin coupling within and between the layers. The subscript $l$ is an index for the top ($l = 1$) and bottom ($l = 2$) layers, 
  $J,J_{\perp}>0$ are the antiferromagnetic couplings between the spins, and 
 $t, t_\perp$ are the intra- and interlayer hopping amplitudes. The restricted fermionic operators, $\tilde{c}^{\dagger}_{i,\sigma} = \hat{c}^{\dagger}_{j,\sigma}(1-\hat{ n }_{i,\bar{\sigma}})$, are constructed so that no double occupancies are allowed. The isotropic spin coupling in Eq. \eqref{Eq:HJ} naturally arises as an effective low-energy description of superexchange processes in the Fermi-Hubbard model at large onsite repulsions $U$, for which $J = 4t^2 / U$ and $J_\perp = 4t_\perp^2 / U$. As this is directly relevant to current experiments using atoms in optical lattices as well as strongly correlated two-dimensional (2D) materials, we will assume these relations in the following such that $J_\perp/J=(t_\perp/t)^2$.

In the absence of holes, the bilayer system undergoes a quantum phase transition at $J_{\perp}/J \simeq 2.5$ \cite{sandvik1994}
corresponding to  $t_\perp / t \simeq 1.6$, from an AFM ordered phase at low $J_{\perp}/J$ to a disordered phase at large $J_{\perp}/J$, in which neighboring spins in the two layers form spin singlets \cite{scalettar1994,kancharla2007a,bouadim2008a}. Deep in the AFM phase, the low energy states can be quantitatively described using spin-wave theory \cite{chubukov1995,vojta1999}, and we will, furthermore, use this framework to qualitatively describe magnetic polarons in the region close to the quantum critical point (QCP).

\subsection{Linear spin-wave theory}
The AFM order defines two sublattices $A$ and $B$ where the spins predominantly point up and down. Using a  Holstein-Primakoff transformation generalized to take the presence of holes into account~\cite{kane1989,martinez1991,liu1991,nyhegn2022}, the mapping for sublattice $A$ is
\begin{align}
	\hat{ S }^{z}_{l,{\mathbf i}} &= (1-\hat{h}^{\dagger}_{l,{\mathbf i}}\hat{h}_{l,{\mathbf i}})/2 - \hat{ s }^{\dagger}_{l,{\mathbf i}}\hat{ s }_{l,{\mathbf i}}, \nonumber \\ 
	\hat S_{l,{\mathbf i}}^{-} &= \hat S^x_{l,{\mathbf i}}-i\hat S^y_{l,{\mathbf i}}=\hat{ s }^{\dagger}_{l,{\mathbf i}} (1-\hat{ s }^{\dagger}_{l,{\mathbf i}}\hat{ s }_{l,{\mathbf i}}-\hat{h}^{\dagger}_{l,{\mathbf i}}\hat{h}_{l,{\mathbf i}})^{1/2}, \nonumber \\
	\tilde{c}_{i,\Dn} &= \hat{h}^{\dagger}_{l,{\mathbf i}}\hat{ s }_{l,{\mathbf i}}, \ \tilde{c}_{i,\Up} = \hat{h}^{\dagger}_{l,{\mathbf i}}(1-\hat{ s }^{\dagger}_{l,{\mathbf i}}\hat{ s }_{l,{\mathbf i}}-\hat{h}^{\dagger}_{l,{\mathbf i}}\hat{h}_{l,{\mathbf i}})^{1/2},
\end{align} 
with $\hat{h}^{\dagger}_{l,{\mathbf i}}$ being a fermionic creation operator of a hole and $\hat{s}^{\dagger}_{l,{\mathbf i}}$ a bosonic creation operator of a spin excitation. Similar expressions apply to sublattice B. Applying the generalized Holstein-Primakoff transformation, keeping only the linear terms, Fourier transforming,  and using the mapping 
\begin{align}
		\begin{bmatrix}
   		\hat{ s }_{1,\bk} \\ 
			\hat{ s }^{\dagger}_{1,-\bk} \\
   		\hat{ s }_{2,\bk} \\ 
			\hat{ s }^{\dagger}_{2,-\bk} 
		\end{bmatrix}
		=
		 \frac{1}{\sqrt{2}}
		\begin{bmatrix}
   			\mathds{1}  &  -\mathds{1} \\
    			\mathds{1}   &  \mathds{1}
		\end{bmatrix}
		\begin{bmatrix}
   			U_{+,\bk} & 0 \\
    			0 &  U_{-,\bk}
		\end{bmatrix}
		\begin{bmatrix}
   		\hat{ b}_{+,\bk} \\ 
			\hat{ b}^{\dagger}_{+,-\bk} \\
   		\hat{ b}_{-,\bk} \\ 
			\hat{ b}^{\dagger}_{-,-\bk} 
		\end{bmatrix},
	\label{Eq:H_J_k}
\end{align}
it has previously been shown \cite{nyhegn2022} that the Hamiltonian transforms into
\begin{equation}
	\hat{ H }_{J} =  \ E_{0} + \sum_{\bk,\mu = \pm}\omega_{\mu, \bk} \hat{ b}^{\dagger}_{\mu,\bk}\hat{ b}_{\mu,\bk}
\end{equation}
and 
\begin{align}
	\hat{ H }_t= &\sum_{\bk,\bp,l} \hat{h}_{l,\bp+\bk}^{\dagger}\hat{h}_{l,\bp}[g_{+}(\bp, \bk)  \hat{ b}^{\dagger}_{+,-\bk}+(-1)^{l}  g_{-}(\bp, \bk) \hat{ b}^{\dagger}_{-,-\bk}] \nonumber \\
	+&\sum_{\bk,\bp} \hat{h}_{2,\bp+\bk}^{\dagger}\hat{h}_{1,\bp}[ f_{+}(\bk) (\hat{ b}_{+,\bk} + \hat{ b}^{\dagger}_{+,-\bk}) \nonumber \\
	  &\phantom{\sum_{\bk,\bp} \hat{h}_{2,\bp+\bk}^{\dagger}\hat{h}_{1,\bp}[\!\!\!\!\!\!}+ f_{-}(\bk) (\hat{ b}_{-,\bk} - \hat{ b}^{\dagger}_{-,-\bk})] + \text{H.c.}
		\label{Eq:HtSlaveFermion}
\end{align}
Here, we only retain the lowest order terms in the operators $\hat{b}^\dagger_{\mu,\bk}$, describing bosonic spin waves in the branch $\mu=\pm$ with dispersions 
\begin{equation}
		\omega_{\pm,\bk} = \frac{1}{2} \sqrt{\left(Jz+J_{\perp}\right)^{2}- \left(Jz\gamma_{\bk} \pm  J_{\perp}  \right)^{2}}.
		\label{Eq:dispersion_rel}
\end{equation} 
The %effective
 intra- and interlayer interaction vertices are %given by  
\begin{align}
	g_{\pm}(\bp,\bk) &= \frac{zt}{\sqrt{2N}}\left[u_{\pm,\bk} \gamma_{\bp + \bk} - v_{\pm,\bk}\gamma_\bp\right],
	\nonumber \\
	f_{\pm}(\bk) &= \frac{t_\perp}{\sqrt{2N}}\left[u_{\pm,\bk} \mp v_{\pm,\bk} \right],
	\label{Eq:f_vertices}
\end{align}
with $z=4$ being the lattice coordinate number, $\gamma_{\bk} = 2(\cos{k_{x}} + \cos{k_{y}} )/z $ the structure factor, and 
 \begin{align}
u_{\pm,\bk} &= \sqrt{\frac{1}{2}\left(\frac{zJ + J_\perp}{2\omega_{\pm,\bk}} + 1\right)} \nonumber \\ 
v_{\pm,\bk} &= {\rm sgn}\left[ z J\gamma_\bk \pm  J_\perp\right]\sqrt{\frac{1}{2}\left(\frac{zJ + J_\perp}{2\omega_{\pm,\bk}} - 1\right)}
\label{Eq:coherence_factors}
\end{align} 
the coherence factors obtained from diagonalizing $\hat{H}_{J}$. The lattice constant is taken to be unity. In the derivation of the quasiparticle wave function, the exact expressions for the interaction vertices are not important, but their symmetries will greatly simplify the expressions. These symmetries  are related to the AFM ordering vector $\bf{Q} = (\pi, \pi)$, and 
for the coherence factors and the spin wave dispersions, we have
\begin{align}
u_{+,\bk+\bQ} &= u_{-,\bk}, \; v_{+,\bk+\bQ} = -v_{-,\bk}, \; \omega_{+,\bk+\bQ} = \omega_{-,\bk},
	\label{Eq:Cohe_symmetry}
\end{align}
whereas the interaction vertices fulfill
\begin{align}
	g_{\pm}(\bp,\bk + \bQ) &= -g_{\mp}(\bp,\bk), \; g_{\pm}(\bp + \bQ,\bk) = - g_{\pm}(\bp,\bk)\nonumber \\
	f_{\pm}(\bk + \bQ) &= f_{\mp}(\bk).
	\label{Eq:interaction_vertex_symmetry}
\end{align}
Equation \eqref{Eq:HtSlaveFermion} constitutes our effective Hamiltonian found within LSWT. For the bilayer Heisenberg model this approximation scheme has been shown to neglect a longitudinal mode \cite{chubukov1995}. To consider this mode, one could instead tend to the bond-operator representation \cite{vojta1999,rademaker2012a, holt2012, holt2013}. That said, the result presented in Sec. \ref{sec.spatial_structure_inter-layer_influence}, when approaching the QCP, reflects the decrease in AFM order which is qualitatively well captured within LSWT, but overestimates the transition point to lie around $t_\perp / t \simeq 3.67$ as can generally be expected from a mean-field theory. With the effective Hamiltonian in place, we turn to the quantum field theory approach applied previously \cite{nyhegn2022} to describe the quasiparticle properties of the emergent magnetic polarons. 

\subsection{Quantum field theory} \label{subsec.quantum_field_theory}
With the effective Hamiltonian in Eq. \eqref{Eq:HtSlaveFermion}, the Matsubara Green's function of a  hole takes on a 
$2\times2$ matrix structure~\cite{nyhegn2022}
\begin{equation}
G_{lm}(\bp, \tau) = - \braket{T_\tau [\hat{h}_{l,\bp}(\tau) \hat{h}^\dagger_{m,\bp}]},
\end{equation}
where $T_\tau$ is the time ordering in imaginary time $\tau$ and  $l,m = 1,2$ denote the two layers. In frequency space, the Dyson equation reads $G(p) = \mathds{1} G_0(p) +  G_0(p) \Sigma(p) G(p)$, where $\Sigma = [\Sigma_{lm}]$ is the self-energy matrix and $p = (\bp, i\omega_p)$ with $i\omega_p$ a fermionic Matsubara frequency. Finally, $G_0(p) = 1/i\omega_p$ is the noninteracting hole Green's function. Due to the layer symmetry, there are only two 
independent matrix elements for the Green's function: $G_{\rd} = G_{11} = G_{22}$ and $G_{o} = G_{12} = G_{21}$ or equivalently $\Sigma_d = \Sigma_{11} = \Sigma_{22}$ and  $\Sigma_o = \Sigma_{12} = \Sigma_{21}$. Using this, it follows from the Dyson equation that
\begin{align}
	G_{\rm d}(p)  = \frac{ G_{0}^{-1}(p) -\Sigma_{\rm d}(p)  }{ \left( G_{0}^{-1}(p) - \Sigma_{\rm d}(p)  \right)^{2}- \left(\Sigma_{\rm o}(p)  \right)^{2} } \nonumber \\
	G_{\rm o}(p)  = \frac{ \Sigma_{\rm o}(p)  }{ \left( G_{0}^{-1}(p) - \Sigma_{\rm d}(p) \right)^{2} - \left(\Sigma_{\rm o}(p)  \right)^{2} }.
	\label{Eq:Prop_D_O}
\end{align} 

\begin{figure}[t!]
	\begin{center}
	\includegraphics{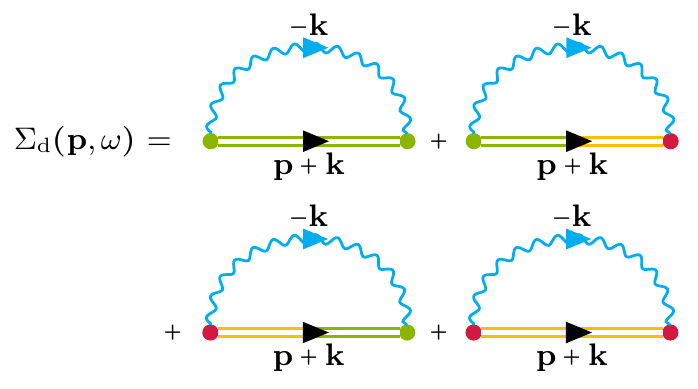}
	\end{center}
	\caption{Feynman diagrams for $\Sigma_{\text{d}}$ within the SCBA. %The color code represents propagation in different layers and whether or not the vertex is associated with inter- or intra-layer hole-spin wave scattering. 
	The green color is associated with layer $l=1$ and yellow with $l=2$, such that the all green and all yellow propagators represent $G_{\rd}$ while the mixed green and yellow propagator represent $G_{\ro}$. The green (yellow) vertex describes a  hole in layer $l=1$ $(l=2)$
interacting with a spin wave and continuing in the same layer. Instead, the red vertex describes interlayer interactions where the hole jumps from one layer to the other. The off-diagonal self-energy $\Sigma_{\rm o}$ is given by similar diagrams ~\cite{nyhegn2022}. }
	\label{Fig:Self}
\end{figure}

In order to calculate the self-energies, we employ the SCBA to sum the class of so-called rainbow diagrams as illustrated in Fig.~\ref{Fig:Self}. This includes an infinite number of spin waves that are absorbed in the same order as they are emitted and turns out to provide an accurate description of hole dynamics in the strong coupling limit $t\gg J$ that naturally arises when $U\gg t$~\cite{nielsen2021}.

Using the vertex symmetries in Eq.~\eqref{Eq:interaction_vertex_symmetry}, we obtain~\cite{nyhegn2022}
 \begin{align}
	\Sigma(\bp,\omega) =& \ \sum_{\bk,\mu} V_{\mu}(\bp,\bk) G(\bp + \bk,\omega - \omega_{\mu,\bk})V^{\dagger}_{\mu}(\bp,\bk) \nonumber \\
	=& \ 2 \sum_{\bk} V_{+}(\bp,\bk) G(\bp + \bk,\omega - \omega_{+,\bk})V^{\dagger}_{+}(\bp,\bk), 
	\label{Eq:Self_Energies}
\end{align}
with the coupling matrix 
 \begin{align}
	V_{\mu}(\bp,\bk) =& \begin{bmatrix} \mu g_{+}(\bp,\bk) &  \mu f_{+}(\bk) \\ f_{+}(\bk) & g_{+}(\bp,\bk) \end{bmatrix}.
	\label{Eq:CouplingMatrix}
\end{align}
We iteratively solve Eqs.~\eqref{Eq:Prop_D_O} and \eqref{Eq:Self_Energies} self-consistently starting from 
 $\Sigma = 0$. This yields the hole Green's function with poles determining the energy of the magnetic polarons, which due to the layer symmetry come in two classes: either symmetric or antisymmetric under layer exchange~\cite{nyhegn2022}.  
  
\begin{figure}[]
	\hspace{-0.7cm}
	\includegraphics[width=1.\columnwidth]{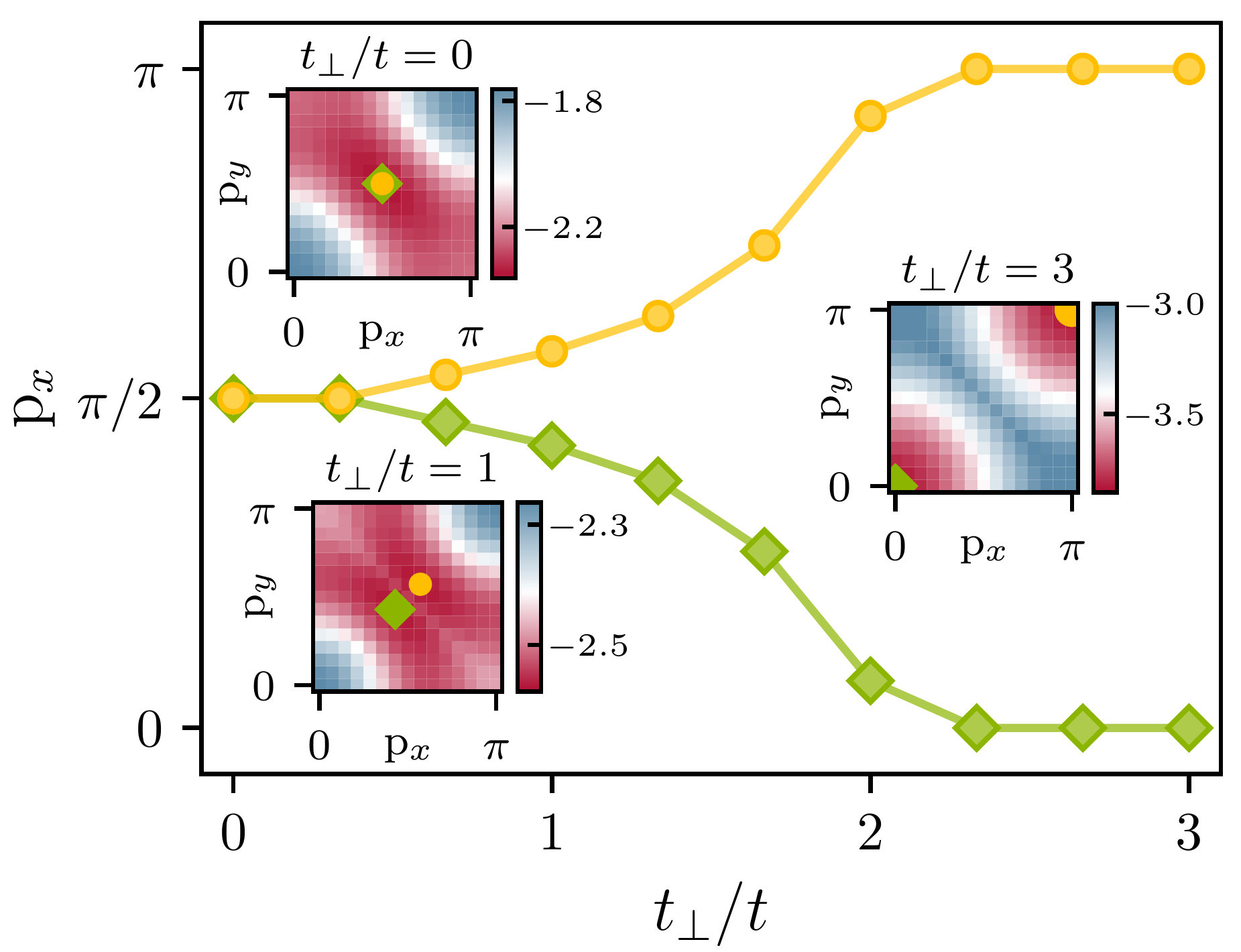}
	\caption{The ground state crystal momentum versus $t_{\perp}/t$ for $J/t = 0.3$. The yellow and green markers correspond to the symmetric and antisymmetric solution respectively and they both represent the degenerate ground state. The insets show the dispersion of the polaron in 
	the first quadrant of the Brillouin zone at written values of $t_{\perp}$ and with the color bars indicating $\omega/t$. The energy minima are indicated by the yellow and green markers which represent the symmetric and antisymmetric solution respectively. Around $t_{\perp}/t\simeq 2.3$, the crystal momentum of the ground state settles at $\bp=\mathbf{0}$ for the antisymmetric solution and $\bp=(\pi,\pi)$ for the symmetric. }
	\label{Fig:GS_movement}
\end{figure}

 In Fig.~\ref{Fig:GS_movement},  we plot the crystal momentum of the lowest energy polaron as a function of $t_\perp/t$. Here and in the rest of the paper we use   $J/t=0.3$. When the layers are decoupled, i.e.\ $t_{\perp}/t = 0$, the magnetic polaron ground state of the system has crystal momentum $\bp=(\pi/2,\pi/2)$. For increasing interlayer hopping, however, Fig.~\ref{Fig:GS_movement} shows that the ground state momentum moves from $\bp=(\pi/2,\pi/2)$ along the diagonal and settles at $\bp={\bf 0}, (\pi, \pi)$ for $t_{\perp}/t\gtrsim 2.3$~\cite{nyhegn2022,vojta1999}. As we will show in Sec. \ref{sec.spatial_structure_inter-layer_influence}, this is essential for understanding the spatial structure of the ground state as the coupling between the layers changes. The insets in Fig.~\ref{Fig:GS_movement} show the energy dispersion of the magnetic polaron in the Brillouin zone for different values of $t_\perp/t$ with the momentum of the ground state indicated. 

\section{Quasiparticle wave function} \label{sec:QP}
While quantities such as the quasiparticle energy and residue are naturally obtained from the field theory described above, it is much less suited to calculate observables related to the real-space structure of the magnetic polaron. This is a serious limitation given the detailed spatial information provided by the optical lattice experiments~\cite{koepsell2019a,ji2021}. Motivated by this, we now generalize a nonperturbative scheme to calculate the many-body wave function of the magnetic polaron within the SCBA from a single layer \cite{reiter1994b,nielsen2021} to the bilayer case. 

The wave function of the magnetic polaron can in general be expanded in the number of spin waves it has excited in the antiferromagnet. This gives  
\begin{align}
\ket{\Psi_\bp} =& \sum_l\Big[ a^{(l)}(\bp) \hat{h}^{\dagger}_{l,\bp}% + a^{(2)}(\bp) \hat{h}^{\dagger}_{2,\bp}
 +\sum_{ \mu_{1}, \bk_{1} }  a^{(l)}_{\mu_{1}}(\bp ,\bk_{1}) \hat{h}^{\dagger}_{l, \bp + \bk_{1}} \hat{b}^{\dagger}_{\mu_{1}, -\bk_{1}} \nonumber \\
  % \!+\! a^{(2)}_{\mu_{1}}(\bp ,\bk_{1})\hat{h}^{\dagger}_{2, \bp + \bk_{1}} \hat{b}^{\dagger}_{\mu_{1}, - \bk_{1}} \big) \nonumber \\
& + \dots \Big] \ket{\AF},
	\label{Eq:Wave}
\end{align}
%which is the natural setup to use within linear spin wave theory. 
where $\ket{\AF}$ is the AFM ground state and 
$\ldots$ represents states with multiple spin waves excited. The coefficient $a^{(l)}_{\{\mu_n\}}(\bp, \{\bk_n\})$ is the amplitude for having $n$ spin waves excited in the branches $\{\mu_n\} = \mu_1, \dots, \mu_n$ with crystal momenta $\{\bk_n\} = -\bk_1, \dots, -\bk_n$. 
The $i$'th spin wave is, hereby, excited into the branch $\mu_i$ with crystal momentum $-\bk_{i}$. The superscript, $l=1,2$ defines the layer
 in which the hole resides. 

\subsection{Recursive determination of amplitudes}
To find explicit expressions for the amplitudes, we use the %time-independent
 Schr\"odinger equation $(\omega - \hat{H})\ket{\Psi_{\bp}}=0$
for the polaron with energy $\omega$. The Hamiltonian given by Eq.~\eqref{Eq:HtSlaveFermion} can only excite and annihilate a single spin wave at a time, any coefficient is, therefore, only coupled to the coefficients for having one more or one less spin wave excited. In addition, when taking matrix elements of the Schr\"odinger equation the SCBA dictates that the order of the spin waves is kept fixed so that only a single Wick contraction for a particular coefficient is maintained. 
This reduces the complexity of the many-body wave function while retaining the possibility of describing a highly correlated state with many spin waves excited. 

With this approach, we derive a set of equations of motion, which at order $n \ge1$ has the matrix form 
\begin{align}
&\left[\omega - \sum_{i = 1}^n \omega_{\mu_i,\bk_i} \right] \ba_{\{\mu_n\}}(\bp, \{\bk_n\})  \nonumber \\
& = V^\dagger_{\mu_n}(\bK_{n-1},\bk_n) \ba_{\{\mu_{n-1}\}}(\bp, \{\bk_{n-1}\}) \nonumber \\
&+\! \sum_{ \substack{\bk_{n+1}\\ \mu_{n+1}} } \! V_{\mu_{n+1}}(\bK_{n}, \bk_{n+1}) \ba_{\{\mu_{n+1}\}}(\bp, \{\bk_{n+1}\}).
\label{eq.EOM}
\end{align}
Here, $\ba = [a^{(1)}, a^{(2)}]$ is a vector containing the wave-function coefficients for the hole in each layer and 
  $\bK_n \equiv \bp + \sum_{i = 1}^n \bk_i$. 
  In close analogy to the single-layer case \cite{reiter1994b,ramsak1998,nielsen2021}, we find that the recursive relation
\begin{align}
&\ba_{\{\mu_{n + 1}\}}(\bp, \{\bk_{n + 1}\}) =  G\big(\bK_{n+1}, \omega - \sum_{i = 1}^{n+1} \omega_{\mu_i,\bk_i}\big) \nonumber \\
& \times  V^{\dagger}_{\mu_{n+1}}(\bK_n, \bk_{n+1}) \ba_{\{\mu_n\}}(\bp, \{\bk_n\})
\label{eq.recursive_relation_1}
\end{align}
self-consistently solves the equations of motion as can straightforwardly be checked by inserting Eq.~\eqref{eq.recursive_relation_1} into Eq.~\eqref{eq.EOM}
(see  Appendix \ref{app.checking_recursive_solution} for details). Equation~\eqref{eq.recursive_relation_1} gives for $n = 0$  the matrix equation 
\begin{align}
&\left[\omega \cdot \mathds{1} - \Sigma(\bp,\omega)\right] \ba(\bp) = 0,
\label{eq.EOM_0}
\end{align}
which shows that $\omega$ must coincide with one of the two poles  $\epsilon^{\pm}_\bp$ of the Green's functions in Eq.~\eqref{Eq:Prop_D_O} in order to have 
a nonzero solution of the coefficients. Thus, the polaron energies obtained from the wave function are the same as from the Green's function. This explicitly 
demonstrates the 
consistency of the two approaches. The  zero-order coefficients for the symmetric or antisymmetric polaron with wave function $\ket{\Psi^{\pm}_{\bp}}$ are 
 $a^{(2)}(\bp) = \pm a^{(1)}(\bp)$ solving $a^{(1)}(\bp) [ \omega - \Sigma_{\rm d}(\bp ,\omega) \mp \Sigma_{\rm o}(\bp ,\omega)] = 0$. 
All higher-order coefficients can now be calculated iteratively in terms of the zero-order coefficients. 

To each order, there will be $2^{n+1}$ terms and an equal number of coefficients. With the vertex symmetries in Eq. \eqref{Eq:interaction_vertex_symmetry}, we can simplify the number of coefficients to a single unique coefficient for each order, which reduces Eq. \eqref{eq.recursive_relation_1} to a scalar relation (see Appendix \ref{app:RedCoeff}). After normalizing the wave function (see Appendix \ref{app.normalization}) the polaron wave function to first order is
\begin{widetext}
	\begin{align}
		\ket{\Psi^{\pm}_{\bp}} = \sqrt{\frac{{Z_{\bp}^{\pm}}}{2}}\left( \hat{h}^{\dagger}_{1,\bp} \pm \hat{h}^{\dagger}_{2,\bp} + \sum_{\bk_{1}} R^{\pm}(\bp ,\bk_{1})  \left[ \left( \hat{h}^{\dagger}_{1, \bp + \bk_1} \pm \hat{h}^{\dagger}_{2, \bp + \bk_1} \right)  \hat{b}^{\dagger}_{+,-\bk_{1}} + \left( \hat{h}^{\dagger}_{1, \bp + \bk_1 + \bQ} \mp \hat{h}^{\dagger}_{2, \bp + \bk_1 + \bQ} \right) \hat{b}^{\dagger}_{-, - \bk_{1} - \bQ}  \right] + ... \right) \ket{\AF},
	\label{Eq:Wave_final}
	\end{align}
\end{widetext}
where 
\begin{align}
	R^{\pm}(\bp ,\{ \bk_{n} \}) =& [ G_{\rm d}( \bK_{n},  \Omega^{ \pm }_{ \{ \bk_{n} \} })  \pm G_{\rm o}( \bK_{n} , \Omega^{ \pm }_{ \{ \bk_{n} \} } )  ]\nonumber\\
	&\times [ g_{+}( \bK_{n-1} ,\bk_{n})  \pm f_{+}( \bk_{n}) ] 
\end{align}	
and 
\begin{equation}	
 \Omega^{ \pm }_{ \{ \bk_{n} \} }(\bp) =\epsilon^{\pm}_\bp - \sum_{i=1}^{n}\omega_{+,\bk_{n}}.
\label{Eq:Omega}
\end{equation}
Here, $Z^{\pm}_{\bp} = [1-\partial_{\omega} \left( \Sigma_{\rd}(\bp ,\omega)  \pm \Sigma_{\ro}(\bp ,\omega)  \right)|_{\omega = \epsilon^{\pm}_{\bp}}]^{-1}$ is the quasiparticle residue of the polaron.

\section{Magnetic dressing cloud} \label{sec.spatial_structure}
Having derived the SCBA expression for the polaron wave function in a bilayer, we now turn to its spatial structure. Specifically, we will 
compute the magnetic dressing cloud, i.e.\ the magnetic frustration around the hole. Following Refs.~\cite{ramsak1998, nielsen2021}, we define the magnetization at 
position $\br+\bd$ in layer $l$ given the hole is at position $\br$  in layer $1$ as
\begin{align}
	M_{l}^{\pm}(\bd,\bp) &= \frac{\braket{\hat{h}^{\dagger}_{1,\br}\hat{h}_{1,\br}\hat{S}^{(z)}_{l,\br + \bd}  }_{\bp}^{\pm}  }{\braket{\hat{h}^{\dagger}_{1,\br}\hat{h}_{1,\br}}^{\pm}_{\bp} \!\braket{\hat{S}^{(z)}_{l,\br + \bd}}_{\bp}^{\pm}} = \frac{1 - 2M_{l}^{(2),\pm}(\bd;\bp,\epsilon^{\pm}_{\bp})}{M_{\AF}},
	\label{Eq:Mag_l}
\end{align}
where   $\braket{\dots}_{\bp}^{\pm} = \bra{\Psi^{\pm}_{\bp}} \dots \ket{\Psi^{\pm}_{\bp}}$ is the average  with respect to the 
symmetric or antisymmetric polaron wave function given by Eq.~\eqref{Eq:Wave_final}. 
The hole-spin correlator is 
\begin{align}
	M_{l}^{(2),\pm}(\bd;\bp,\epsilon^{\pm}_{\bp}) &=  2N\braket{\hat{h}^{\dagger}_{1,\br}\hat{h}_{1,\br}\hat{s}^{\dagger}_{l,\br+\bd }\hat{s}_{l,\br+\bd }}_{\bp}^{\pm} \nonumber \\
	&=  2N\braket{\hat{h}^{\dagger}_{l,\br}\hat{h}_{l,\br}\hat{s}^{\dagger}_{1,\br+\bd }\hat{s}_{1,\br+\bd }}_{\bp}^{\pm}
	\label{Eq:Mag_l_2}
\end{align}
where we have used that the hole is evenly spread out over the entire lattice for crystal momentum states, $\braket{\hat{h}^{\dagger}_{1,\br} \hat{h}_{1,\br}}_{\bp,\pm} = 1/(2N)$. This also means that the hole has negligible impact on the average local magnetization, $\braket{\hat{S}^{z}_{m,\br + \bd}}_{\bp}^{\pm} = \bra{\AF} \hat{S}^{z} \ket{\AF} + \mathcal{O}(1/N)$.
Spin wave theory gives  $2|\bra{\AF} \hat{S}^{z} \ket{\AF}| \equiv M_{\AF} = 1-2M_{\rm fl}$, where $M_{\rm fl} = \sum_{\mathbf{k}} \left( v_{\mathbf{k},+}^{2} + v_{\mathbf{k},-}^{2} \right)/(2N)$ describes the reduction of magnetic order due to quantum fluctuations in the antiferromagnetic ground state. Additionally, we have used the layer symmetry of the system to swap the indices of the holes and spin excitations. Using Eq.~\eqref{Eq:H_J_k}, we can write 
\begin{align}
	M_{l}^{(2),\pm}(\bd;\bp,\epsilon^{\pm}_{\bp}) = M_{\text{fl}} + B^{\pm}_{l,\bp}(\bd) + C^{\pm}_{l,\bp}(\bd).
\end{align}
where the $B$ and $C$ functions describe the frustration induced by the hole. Their explicit expressions read
\begin{widetext}
\begin{align}
	B^{\pm}_{l,\bp}(\bd) &= \frac{1}{N}\sum_{ \bq_{1},\bq_{2} } e^{i(\bq_{1}-\bq_{2})\cdot \bd } B^{\pm}_{l}(\bq_{1},\bq_{2};\bp,\epsilon^{\pm}_{\bp}), \nonumber \\ 
	B^{\pm}_{l}(\bq_{1},\bq_{2};\bp,\omega) &=  \sum_{\bk,\mu_{1},\mu_{2}} b_{\mu_{1},\mu_{2}}(\bq_{1},\bq_{2})  \braket{\hat{h}^{\dagger}_{l,\bk + \bq_{1} - \bq_{2} }\hat{h}_{l,\bk }\hat{b}^{\dagger}_{\mu_{1},-\bq_{1} }\hat{b}_{\mu_{2},-\bq_{2} }}_{\bp,\omega}^{\pm},
	\label{Eq:B_series}
\end{align}
and 
\begin{align} 
	C^{\pm}_{l,\bp}(\bd) &= -\frac{1}{2N}\sum_{ \bq_{1},\bq_{2} } e^{i(\bq_{1}+\bq_{2})\cdot \bd } \cdot C^{\pm,l}(\bq_{1},\bq_{2};\bp,\epsilon^{\pm}_{\bp})
	 + \text{H.c.}, \nonumber \\ 
	 C^{\pm}_{l}(\bq_{1} ,\bq_{2};\bp,\omega) &= \sum_{\bk,\mu_{1},\mu_{2}} c_{\mu_{1},\mu_{2}}(\bq_{1},\bq_{2}) \braket{\hat{h}^{\dagger}_{l,\bk + \bq_{1} - \bq_{2} }\hat{h}_{l,\bk }\hat{b}_{\mu_{1},-\bq_{1} }\hat{b}_{\mu_{2},-\bq_{2} }}_{\bp,\omega}^{\pm}.
	\label{Eq:C_series} 
\end{align}
\end{widetext}
Here  $\langle\ldots\rangle_{\bp,\omega}^{\pm}$ means the average with respect to the polaron wave function $|\Psi_{\mathbf p}^{\pm}\rangle$ given by Eq.~\eqref{Eq:Wave_final} but with the energy 
$\epsilon^{\pm}_{\bp}$ replaced by $\omega$ when calculating the coefficients. The vertex functions $b_{\mu_{1},\mu_{2}}(\bq_{1},\bq_{2}) = \mu_{1}\mu_{2}( v_{\mu_{1},\bq_{1}}v_{\mu_{2},\bq_{2}} + u_{\mu_{1},\bq_{1}}u_{\mu_{2},\bq_{2}})$ and $c_{\mu_{1},\mu_{2}}(\bq_{1},\bq_{2}) = \mu_{1}\mu_{2}(u_{\mu_{1},\bq_{1}}v_{\mu_{2},\bq_{2}} + v_{\mu_{1},\bq_{1}}u_{\mu_{2},\bq_{2}})$ arise due to the transformation to the spin wave operators $\hat{b}$. We have $M_{l}^{\pm}= 1$ for 
$B^{\pm}_{l,\bp} = C^{\pm}_{l,\bp}=0 $, meaning that the magnetization is that of the underlying antiferromagnet in the absence of holes. It follows that $M_{l}^{\pm} < 1$ reflects a hole-induced suppression of AFM order while $M_{l}^{\pm} > 1$ gives an increased order. 
 % The magnetization can also turn negative, showing that the spin has partially flipped compared to the system in absence of holes. The possible values of $M_{l}^{\pm}$ are dictated by $M_{\AF}$ through $|M_{l}^{\pm}| \leq 1/M_{\AF}$. 

Calculating the magnetization around the hole is now reduced to computing the  $B^{\pm}_{l,\bp}$ and $C^{\pm}_{l,\bp}$ functions in Eqs.~\eqref{Eq:B_series} and \eqref{Eq:C_series}. For strong coupling $t/J>1$, it is 
important to do this including terms in the polaron wave function $|\Psi_{\mathbf p}^{\pm}\rangle$ with an arbitrary number of spin waves. This was first achieved recently for a single layer in Ref.~\cite{nielsen2021}, and here we generalize this to the bilayer case at hand. As detailed in Appendixes \ref{app:BSeries} and \ref{app:CSeries}, using a diagrammatic approach we derive self-consistency equations for $B^{\pm}_{l,\bp}$ and $C^{\pm}_{l,\bp}$, which we then solve numerically. This allows us to include terms in the wave function with an arbitrary number of spin waves, which is crucial in the strong coupling regime.  

As described in Sec.~\ref{sec:QP}, the bilayer supports magnetic polaron eigenstates that are either symmetric or antisymmetric with respect to layer exchange. It follows from the AFM symmetries in Eqs.~\eqref{Eq:Cohe_symmetry} and \eqref{Eq:interaction_vertex_symmetry} that the associated magnetization clouds are related by the mapping (see Appendix \ref{app:sym})
 \begin{align}
 	M_{l}^{-}(\bd,\bp + \bQ) &= M_{l}^{+}(\bd,\bp).
	\label{Eq:sym}
 \end{align}
In the following, we, therefore, focus only on antisymmetric eigenstates. Also, we can in analogy with the single-layer case use time-reversal and inversion symmetry to show that the dressing cloud around the polaron is $C_2$ symmetric, even when the momentum of the polaron is nonzero. Indeed, we have $M_{l}^{\pm}(\bd,\bp) = M_{l}^{\pm}(\bd,-\bp)=M_{l}^{\pm}(-\bd,\bp)$ where the first equality follows from time-reversal symmetry and the second from 
inversion symmetry~\cite{nielsen2021}. It follows from these symmetries that all degenerate polaron states have the same dressing cloud.

\section{Results} \label{sec.spatial_structure_inter-layer_influence}
We now discuss our numerical results for the dressing cloud of magnetic frustration around the hole. Compared to the single layer case discussed in Ref.~\cite{nielsen2021}, the bilayer system has a qualitatively new feature in the 
sense that it undergoes a quantum phase transition to a disordered state with increasing interlayer coupling~\cite{scalettar1994,kancharla2007a,bouadim2008a}. We, therefore, focus on how this interlayer coupling affects the magnetic dressing cloud of the polaron. The calculations are performed on a $16\times16$ lattice with periodic boundary conditions by first solving the self-consistent equations for the Green's functions, after which the $B^{\pm}_{l,\bp}$ and $C^{\pm}_{l\bp}$ series are calculated. 

\subsection{Dressing cloud for fixed momentum}\label{Fixedp}
\begin{figure}[t!]
\begin{center}
	\includegraphics[width=1.0\columnwidth]{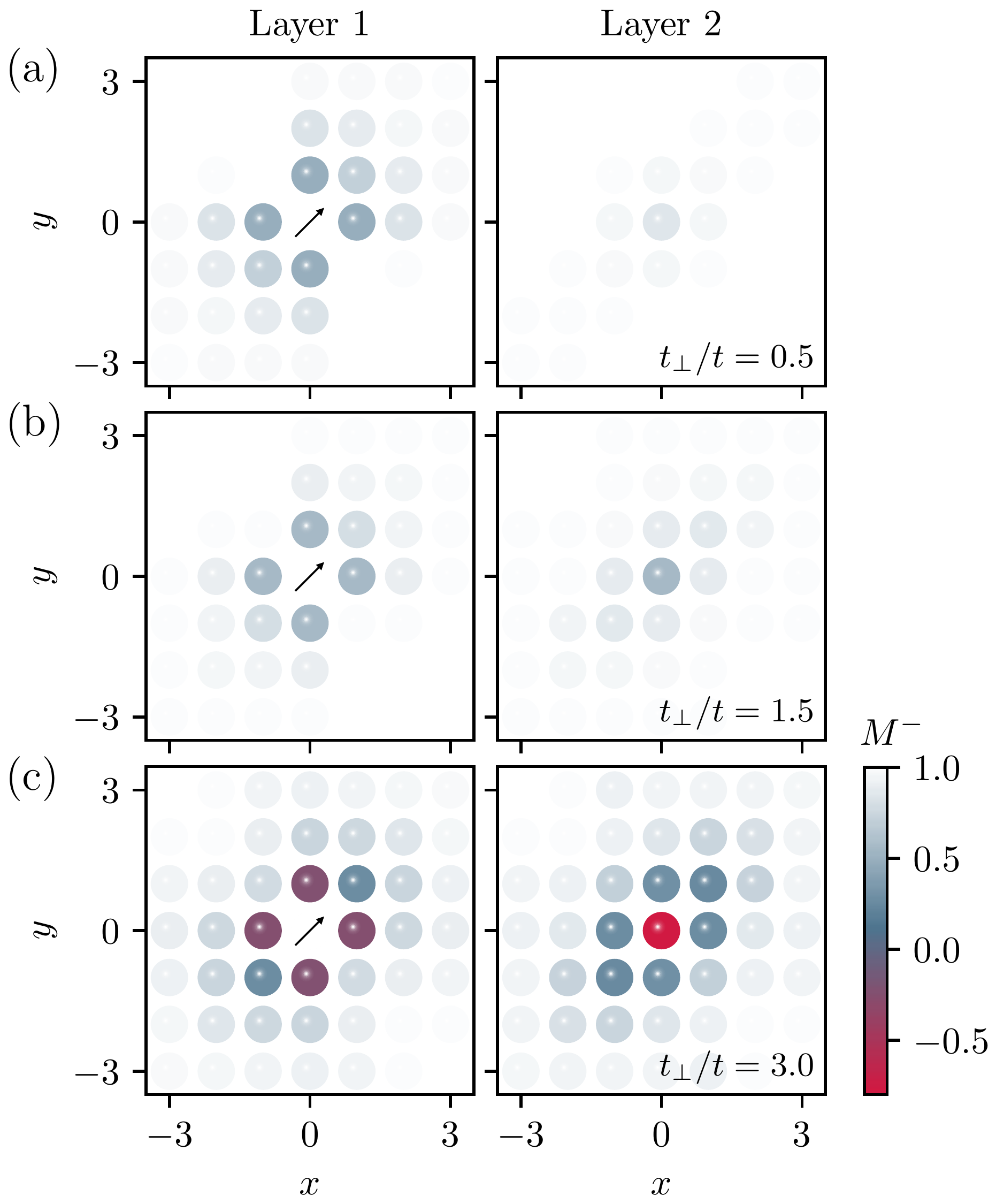}
	\end{center}
	\caption{The dressing cloud of magnetic frustration  in layer 1 (left) and layer 2 (right) around the hole in layer 1 for the polaron with 
	 crystal momentum $\bp =(\pi / 2, \pi / 2)$ indicated by the arrow. The cloud has a $C_{2v}$ symmetry in both layers with the spins primarily frustrated parallel to the crystal momentum. For $t_{\perp}/t = 0.5,1.5$ the frustration cloud is clearly more prominent in the same layer as the hole. Instead for $t_{\perp}/t = 3$, we see clouds of similar sizes in the two layers.}
	\label{Fig:MagFrus_1}
\end{figure}
Figure  \ref{Fig:MagFrus_1} shows the magnetization $M_{l}^{-}(\bd,\bp)$ at sites separated by the in-layer vector ${\mathbf d} = (x,y)$ from the hole for the antisymmetric polaron. We take  momentum ${\mathbf p}=(\pi/2,\pi/2)$, which is the ground state for $t_\perp/t\lesssim0.6$ (see Fig.~\ref{Fig:GS_movement}). The left column shows the magnetization in layer $1$ where the hole is located, and the right column shows the magnetization in layer $2$. Different rows correspond to different values of $t_\perp/t$. The value  $M_{l}^{-}(\bd,\bp)=1$ corresponds to a magnetization equal to that of the AFM state without the hole. Figure \ref{Fig:MagFrus_1} clearly shows an elongated shape of frustration in both layers and for all values of $t_\perp/t$ with the dressing cloud primarily spread out parallel to the diagonal $x = y$. This is analogous to what was previously shown to be the case for crystal momenta along the magnetic Brillouin zone (MBZ) boundary $|p_x|+|p_y|=\pi$ for a single layer. We also see that, in addition to the $C_2$ symmetry, the dressing cloud has mirror symmetries along the diagonals. One can in fact show that for momenta along the MBZ boundary, the magnetization cloud has the symmetries of the group $C_{2v}$~\cite{nielsen2021}. We, furthermore, see that with increasing $t_\perp$ the dressing cloud steadily increases in spatial size and magnitude in layer $2$. Contrary to this, the dressing cloud in layer $1$ where the hole resides is initially slightly decreasing with increasing $t_\perp$. It eventually starts increasing for larger values of $t_\perp$ and the dressing cloud has a similar size in the two layers for $t_{\perp}/t = 3$, where the hole has even reversed the sign of the magnetic order at neighboring lattice sites. 
\begin{figure}
	\begin{center}
	\hspace{-0.5cm}
	\includegraphics[width=1.0\columnwidth]{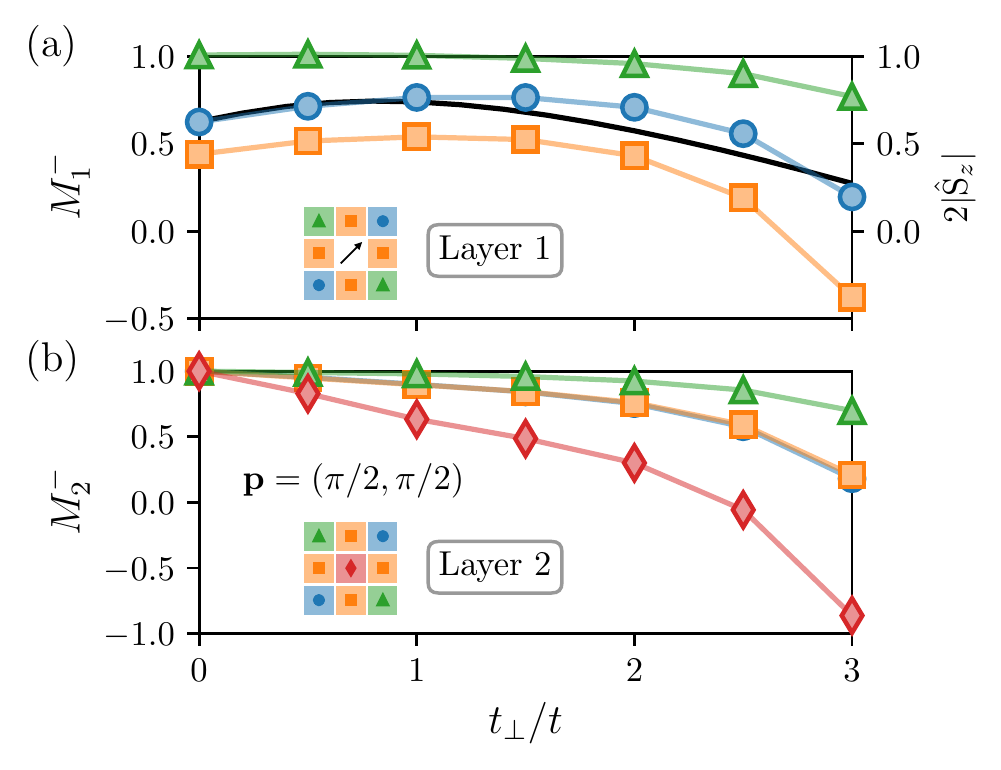}
	\end{center}
	\caption{The magnetization in layer (a) 1 and (b) layer 2 in the vicinity of a hole in layer 1 as a function of interlayer hopping 
	 for $\bp =(\pi/2,\pi/2)$. Deviations from $M = 1$ signals frustration. (a) In layer 1, we see a nonmonotonic behavior as $t_{\perp}/t$ is increased. This can be understood through the strength of the AFM order, which is shown as a black line. As the order increases the cloud diminishes in size before expanding as the order decreases. (b) In layer 2 on the other hand, the magnetic frustration monotonically increases.}
	\label{Fig:MagFrus_t_perp}
\end{figure}
To analyze this behavior further, we plot in  Fig.~\ref{Fig:MagFrus_t_perp} the magnetization in the vicinity of the hole as a function of $t_\perp/t$. 
We also plot  the  magnetization $2|\bra{\AF} \hat{S}^{z} \ket{\AF}|$ in the absence of holes,  which 
exhibits a nonmonotonic behavior: It  initially increases with  $t_\perp$ since  the Fermi-Hubbard relation $J_\perp\propto t_\perp^2$ makes the AFM ordering across the 
planes stronger, reaching a maximum at $t_\perp/t\simeq 0.82$ after which it decreases as the quantum phase transition to the disordered state is 
approached~\cite{nyhegn2022}. This nonmonotonic behavior of the magnetic order is reflected in the size and magnitude of the dressing cloud in 
layer 1, which initially decreases with $t_\perp$,  i.e. $M$ increases towards $1$ as can be seen in Fig.~\ref{Fig:MagFrus_t_perp}, because the increased size of the order parameter reduces the mobility of the hole. 
On the other hand, when $t_\perp > 0.82t$ and the local AF order starts decreasing, it becomes easier for the hole to delocalize which boosts the magnetic frustration around the hole (see Fig.~\ref{Fig:MagFrus_t_perp}). For small values of $t_\perp/t$, the second layer is essentially not affected by the presence of a hole in layer 1, and the magnetization is hardly changed from its background value, $M^-_2 = 1$. However, as the coupling between the layers is increased, frustration appears here as well. This happens both because the hole can now make virtual hops to the second layer via $t_\perp$, and because the interlayer spin coupling $J_\perp \propto t_\perp^2$ favors the spins in the second layer to antialign with the spins in layer 1. Again we see from Fig.~\ref{Fig:MagFrus_t_perp} that for large values of $t_\perp$ the presence of the hole even inverts the sign of the magnetic order. For values other than $J / t = 0.3$, the situation is qualitatively the same. Quantitatively, lower values of $J / t < 0.3$ mean a lowered frustration cost and result in an increase in frustration, and vice versa for  $J / t > 0.3$. This is detailed in Appendix \ref{App:J_t}.

\subsection{Dressing cloud of ground state}
\begin{figure}
	\begin{center}
	\includegraphics[width=1.0\columnwidth]{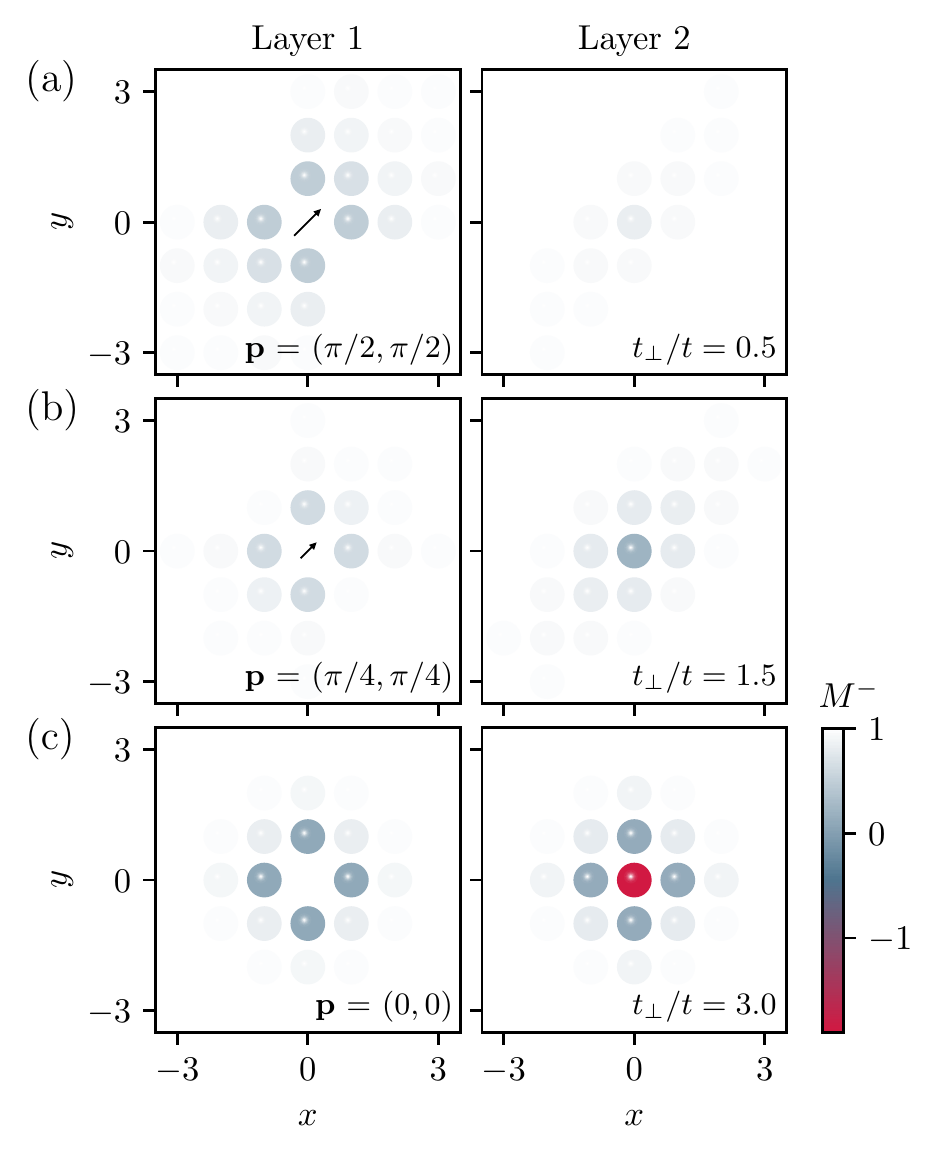}
	\end{center}
	\caption{The magnetization cloud in layer 1 (left) and layer 2 (right) in the vicinity of a hole in layer 1 for the ground state of the system. The ground state momentum changes from (a) $\bp = (\pi/2,\pi/2)$, over (b) $\bp = (\pi/4,\pi/4)$, to  (c) $\bp = {\bf 0}$ as illustrated by the black arrows.}
	\label{Fig:MagFrus_2.5}
\end{figure}

As we discussed in Sec.~\ref{subsec.quantum_field_theory}, the momentum of the ground state polaron smoothly
changes from $\bp=(\pi/2,\pi/2)$ at $t_\perp=0$ to $\bp = {\bf 0}$ [or $\bp = (\pi,\pi)$] as the QCP is approached with increasing $t_\perp$ (see Fig.~\ref{Fig:GS_movement}~\cite{nyhegn2022,vojta1999}). Figure \ref{Fig:MagFrus_2.5} plots the magnetization in the vicinity of the hole as a function of $t/t_\perp$ for the ground state, i.e.\ the crystal momentum varies according to Fig.~\ref{Fig:GS_movement}. In addition to the increase of the dressing cloud with increasing $t_\perp$ also seen for the case of the fixed momentum discussed in Sec.~\ref{Fixedp}, Fig.~\ref{Fig:MagFrus_2.5} shows that the \emph{symmetry} of the dressing cloud changes. The reason for this change is that the crystal momentum of the ground state varies with $t_\perp$. For small $t_\perp/t$, the momentum $\bp=(\pi/2,\pi/2)$ is at the magnetic Brillouin zone boundary so that the dressing cloud has $C_{2v}$ symmetry \cite{nielsen2021}. As seen in Fig.~\ref{Fig:GS_movement}, the momentum of the ground state decreases with $t_\perp$, becoming zero for $t_\perp/t\gtrsim2.3$. At this point, the dressing cloud recovers the full $C_{4v}$ symmetry of the AFM order. It follows that the symmetry of the magnetic polaron undergoes the transition $C_{2v}\rightarrow C_{4v}$ with increasing $t_\perp/t$. 

\begin{figure}
	\begin{center}
	\hspace{-1.5cm}
	\includegraphics[width=0.4\textwidth]{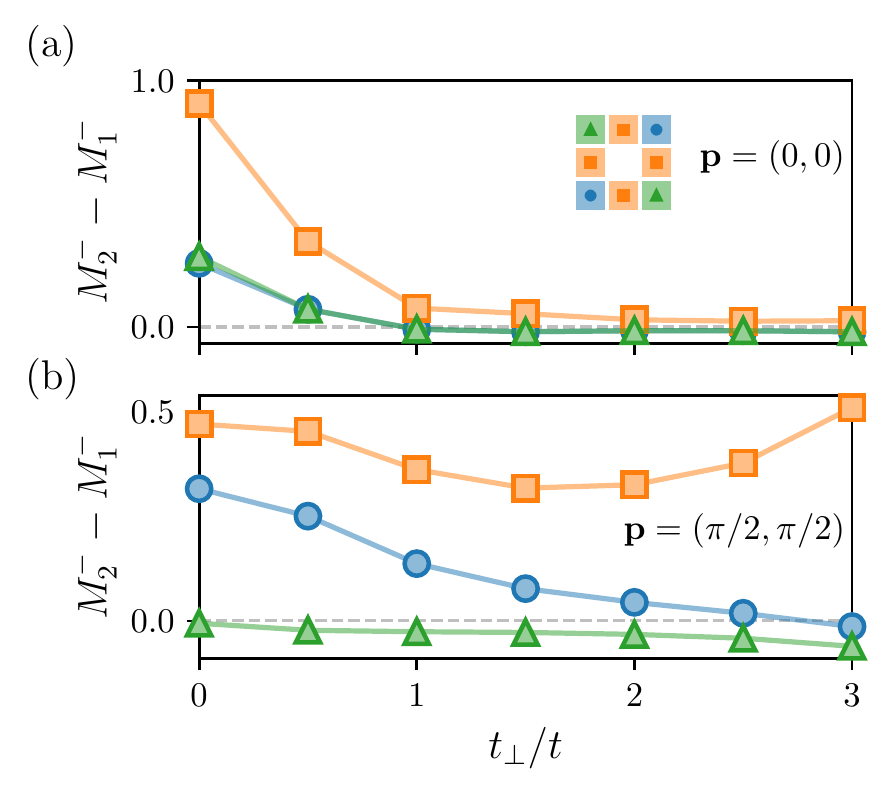}
	\end{center}
	\caption{Difference $M^{-}_1 - M^{-}_2$ in magnetization between the layers versus the interlayer hopping $t_{\perp}/t$. (a) For ${\bf p} = {\bf 0}$, the magnetization rapidly becomes identical in the two layers. (b) For $\bp = (\pi/2,\pi/2)$, they remain different for the shown values of $t_{\perp}/t$ but will eventually become identical (Appendix \ref{app:BSeries}).}
	\label{Fig:DeltaMag}
\end{figure}

Figure \ref{Fig:MagFrus_2.5} also shows that the dressing cloud in layer 2 closely mimics that in layer 1 when $t_\perp$ is large. 
 To analyze  this quantitatively, we plot in Fig.~\ref{Fig:DeltaMag}  the difference $M_1^-(\br)-M_2^-(\br)$ between the magnetization  in the two 
 layers for $\bp = (\pi/2,\pi/2)$ and $\bp = {\bf 0}$ for a few 
 selected sites as a function of $t_{\perp}/t$. This clearly shows that the  difference vanishes rapidly with increasing $t_\perp$ for the polaron with momentum 
  $\bp = {\bf 0}$  [Fig. \ref{Fig:DeltaMag}(a)]. The reason is that the spins in the dressing cloud in the two layers antialign for strong interlayer coupling making them approximately mirror images of each other. 
   Notably, this antialignment of the spins inside the dressing clouds in the two layers is much less pronounced for the polaron with momentum $\bp = (\pi/2,\pi/2)$, and this yields spatial insight into why the ground state momentum changes from  $\bp = (\pi/2,\pi/2)$ to $\bp = {\bf 0}$ for increasing $t_\perp$. In fact, for $t_\perp \gg t$, it comes with a high magnetic energy cost scaling as $J_\perp \propto t_\perp^2$ if spins do not antialign between the layers, explaining why the $\bp = (\pi/2,\pi/2)$ polaron has higher energy than the $\bp = {\bf 0}$ polaron in this regime. In Appendix \ref{App:Decoupled}, we elaborate on how the interlayer hopping and interlayer spin-spin individually promote this mirroring effect of the ground state by taking $J_{\perp}\neq 4t^{2}_{\perp}/U$. Here, we see that the hopping primarily causes the mirroring effect, while the spin-spin interaction penalizes states not having AFM order between the layers. As a result, both couplings are needed to ensure that the ground state features mirrored layers for increasing $t_\perp / t$.  

\begin{figure}
	\begin{center}
	\includegraphics[width=0.4\textwidth]{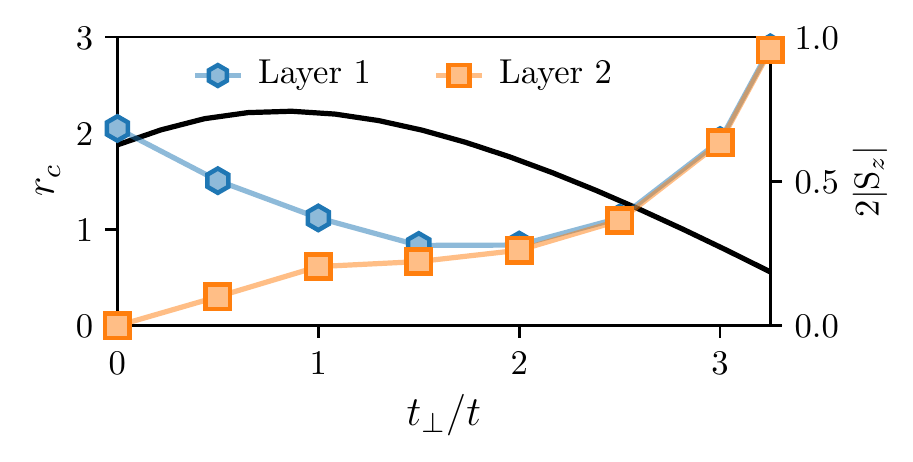} 
	\end{center}
	\caption{The size of the ground state dressing cloud in the two layers with the hole residing in layer $1$ as a function of $t_{\perp}/t$. The size of the magnetic order parameter in the absence of the hole is shown as a black line. }
	\label{Fig:Dist}
\end{figure}

Finally, we plot in Fig.~\ref{Fig:Dist} the size of the magnetic dressing cloud of the ground state in the two layers defined as 
\begin{equation}
r_{c}^{l}(\bp) = \sqrt{\sum_{\bd}|\bd|^{2} \cdot |M^{-}_{l}(\bd,\bp)-1|}
\end{equation}
as a function of $t_\perp/t$. We subtract $1$ from the frustration since $M^{-}_{l}(\bd,\bp)-1=0$ corresponds to no frustration. This quantifies how the dressing cloud reflects the size of the order parameter of the environment and, in particular, how it grows as the phase transition to the disordered state, occurring at $t_\perp/t\simeq 3.67$ within LSWT,  is approached.  We believe this increase is a reliable result even though spin-wave theory cannot describe the region close to the QCP in a quantitatively accurate way. The reason is that the increase is caused by the increase in the hole mobility, which in turn is due to the decrease in the AF order as the QCP is approached. This is a robust physical effect and not an artifact of the theoretical approximations. In Refs. \cite{holt2012,holt2013} they argue for the same expansion of the dressing cloud when the QCP is approached from the disordered phase, supporting our results. The same effect is shown in Fig.~\ref{Fig:Front}, which clearly shows how the magnetic dressing cloud expands as  $t_\perp$ increases and the phase transition to the disordered state is approached. One can also see how the dressing clouds in the two layers become mirror images of each other with their spins antiparallel for large $t_\perp/t$.
 %%%%%%%%%%%%%%%%%%%%%%%%%%%%%%%%%%%%%%%
%%%%%%%%%%%%%%%%%%%%%%%%%%%%%%%%%%%%%%%

\section{Conclusions}\label{conclusions}
Inspired by the impressive ability to experimentally probe the spatial properties of strongly correlated fermions with single-site resolution, not only in single-layer but also bilayer optical lattices~\cite{gall2021}, we developed a nonperturbative scheme for calculating the polaron wave function for a bilayer antiferromagnet. The scheme is the wave function version of the SCBA~\cite{reiter1994b,ramsak1998,nielsen2021} and allows one to include an infinite number of spin waves, which is crucial for describing the strongly correlated regime. 
With the developed wave function formalism, it is possible to calculate the spatial structure of the bilayer magnetic polarons via spin-hole correlators. In particular, we mapped out the associated dressing cloud in the vicinity of a hole and investigated how this was influenced by the interlayer coupling strength. Here, we observed the cloud to mimic the nonmonotonic behavior of the antiferromagnetic order, where it first contracts due to an increase in the AFM order, whereafter it starts to expand as the quantum phase transition to the disordered phase is approached. 
We, furthermore, found that the symmetry of the ground state dressing cloud undergoes the transition  $C_{2v}\rightarrow C_{4v}$ with increasing $t_\perp/t$. This happens because the ground state momentum smoothly changes along the diagonal from $\bp = (\pi/2,\pi/2)$ for small $t_\perp/t$  to $\bp = \bf{0}$ for large $t_\perp/t$. In the latter limit, we find that the strong interlayer coupling makes the spins in the two layers antiparallel in the dressing cloud so that they become mirror images of each other. These effects should be amenable to direct observation in quantum simulation experiments. Here, however, it is important to note that one must project onto a single crystal momentum eigenstate to observe something else than the full $C_{4v}$ symmetry of the underlying system.

Our scheme for nonperturbatively calculating the polaron wave function in a bilayer geometry within SCBA offers a tool that can help to elucidate the behavior of dopants in antiferromagnetic environments. For a single layer, the wave function has been used to investigate the nonequilibrium physics associated with the creation of the magnetic polaron \cite{nielsen2022}. The similarities between the expressions for the single-layer and bilayer system suggest that it is possible to probe the same dynamics for a bilayer system with the here-derived wave function. It also offers the possibility of a detailed comparison with novel ultracold-atom experiments \cite{gall2021}. 

In the future, we hope to explore extensions to finite temperatures, as well as comparisons with sophisticated numerical methods, using, e.g., density matrix renormalization techniques \cite{white1997,maier2011}. This will allow a more quantitative description as the quantum phase transition from the AFM ordered state at low interlayer coupling to the disordered phase at large interlayer coupling is crossed. 

\begin{acknowledgments}
This work has been supported by the Danish National Research Foundation through the Center of Excellence “CCQ” (Grant Agreement No.: DNRF156), as well as by the Carlsberg Foundation through a Carlsberg Internationalisation Fellowship. 
\end{acknowledgments}

\clearpage

%%%%%%%%%%%%%%%%%%%%%%%%%%%%%%%%%%%%%%%%%%%%%%%%%%%%%%%%%%%%%%%%%%%%%%%%%%%%%%%%%%%%%%%%%%%%%%%%%%%%%%%%%%
%%%%%%%%%%%%%%%%%%%%%%%%%%%%							Appendix							    %%%%%%%%%%%%%%%%%%%%%%%%%%%%%%%%%%%%%%%
%%%%%%%%%%%%%%%%%%%%%%%%%%%%%%%%%%%%%%%%%%%%%%%%%%%%%%%%%%%%%%%%%%%%%%%%%%%%%%%%%%%%%%%%%%%%%%%%%%%%%%%%%%

\appendix

\section{Recursion relation}  \label{app.checking_recursive_solution}
In this appendix, we will explicitly show that the recursion relation presented in Eq. \eqref{eq.recursive_relation_1} self-consistently solves Eq. \eqref{eq.EOM}. Inserting the recursion relation into Eq. \eqref{eq.EOM}, we find
\begin{align}
	&\left[\left( \omega - \sum_{i = 1}^n \omega_{\mu_i,\bk_i}\right) \identity -  \Sigma\big(\bK_{n}, \omega - \sum_{i = 1}^{n} \omega_{\mu_i,\bk_i}\big)  \right] \ba_{\{\mu_n\}}(\bp, \{\bk_n\}) \nonumber \\
	&=V^\dagger_{\mu_n}(\bK_{n-1},\bk_n) \ba_{\{\mu_{n-1}\}}(\bp, \{\bk_{n-1}\}),
	\label{Eq:Rec_self}
\end{align}
where Eq. \eqref{Eq:Self_Energies} has been used. Looking at the Green's functions in Eq. \eqref{Eq:Prop_D_O}, we find
\begin{align}
	G(\bp,\omega)^{-1} =& \begin{bmatrix} \omega-\Sigma_{\rm d}(\bp,\omega)&  -\Sigma_{\rm o}(\bp,\omega) \\ -\Sigma_{\rm o}(\bp,\omega) & \omega-\Sigma_{\rm d}(\bp,\omega) \end{bmatrix},
	\label{Eq:CouplingMatrix}
\end{align}
which corresponds to the parentheses on the left-hand side in Eq. \eqref{Eq:Rec_self}. Inserting this, we retrieve
\begin{align}
	&\ba_{\{\mu_n\}}(\bp, \{\bk_n\})  =  G\big(\bK_{n}, \omega - \sum_{i = 1}^{n} \omega_{\mu_i,\bk_i}\big)\nonumber \\
	&\times V^\dagger_{\mu_n}(\bK_{n-1},\bk_n) \ba_{\{\mu_{n-1}\}}(\bp, \{\bk_{n-1}\}),
\end{align}
showing the recursion relation to be self-consistent.

\section{Normalization} \label{app.normalization}
In this Appendix, we show how the normalization of the wave function links the lowest-order amplitude to the quasiparticle residue. We have,
\begin{align}
	1 = \ &\bracket{ \Psi^{\pm}_{ \bp } }{ \Psi^{\pm}_{ \bp } } = 2 \left( a^{\pm}_{0}(\bp) \right)^{2} + 2^{2}\sum_{ \bk_{1} } \left( a^{\pm}_{1}( \bp ,\bk_{1} ) \right)^{2}  +   \nonumber \\ 
	&2^{3}\sum_{ \bk_{1} , \bk_{2}  } \left( a^{\pm}_{2}(\bp, \{\bk_{2}\}) \right)^{2}  + ...,
\end{align}
where the factor $2^{n+1}$ in front comes from the number of different states possible to each order. Each amplitude is related to the previous order through Eq. \eqref{Eq:RecRel}. Inserting this yields
\begin{align}
	&\bracket{ \Psi^{\pm}_{ \bp } }{ \Psi^{\pm}_{ \bp } } = 2 \left( a^{\pm}_{0}(\bp) \right)^{2} \left( 1 +  2 \sum_{ \bk_{1} } \left[ R^{\pm}(\bp ,\{ \bk_{1} \}) \right]^{2}  +  \right. \nonumber \\ \left.
	\right. & \left. 2^{2}\sum_{ \bk_{1} , \bk_{2}  } \left[ R^{\pm}(\bp , \{ \bk_{2} \})  \right]^{2} \left[ R^{\pm}(\bp ,\{ \bk_{1} \})  \right]^{2}  + ... \right).
	\label{Eq:Norm2}
\end{align}
Using the expressions for the Green's functions in Eq. \eqref{Eq:Prop_D_O} and the self-energies in Eq. \eqref{Eq:Self_Energies}, we obtain 
\begin{align}
1 -	&\partial_{\omega} \left[ \Sigma_{\rd}(\bp, \omega)  \pm \Sigma_{\ro}(\bp, \omega)\right] = 1 + 2\sum_{\bk_{1}}  \left[ R^{\pm}(\bp ,\{ \bk_{1} \}) \right]^{2} \nonumber \\ 
	&\left(1 - \partial_{\omega} \left[ \Sigma_{\rd}( \bk_{1}, \omega - \omega_{+,\bk})  \pm \Sigma_{\ro}( \bk_{1}, \omega - \omega_{+,\bk}) \right] \right).
\end{align}
Evaluating this at $\omega = \varepsilon^{\pm}_\bp$ and comparing it to Eq. \eqref{Eq:Norm2}, we see that they are identical up to the factor of $2 ( a^{\pm}_{0}(\bp))^{2}$. Therefore,
\begin{align}
	a^{\pm}_{0}(\bp) &= \frac{1}{\sqrt{2}} \frac{1}{\sqrt{1-\partial_{\omega} \left( \Sigma_{\rd}(\bp ,\omega)  \pm \Sigma_{\ro}(\bp ,\omega)  \right)|_{\omega = \epsilon^{\pm}_{\bp}} }},
	\label{Eq:Norm3}
\end{align}
Finally, we relate this to the quasiparticle residue. To make the residue coincide with that for a single layer in the limit of $t_{\perp}/t=0$, we define this as half the residue of the poles in the Green's function whereby
\begin{align}
	Z^{\pm}_\bp /2 = {\rm Res}\left( G_{\rd},\epsilon^{\pm}_\bp \right) &= \left(a^{\pm}_{0}(\bp)\right)^{2} \nonumber \\
	Z^{\pm}_\bp /2 = {\rm Res}\left( G_{\ro},\epsilon^{\pm}_\bp \right) &= \pm \left(a^{\pm}_{0}(\bp)\right)^{2}.
\end{align} 
Consequently, the quasiparticle residue, hereby, describes the total probability of finding the magnetic polaron as a bare hole
\begin{align}
	|\bra{ \Psi^{\pm}_{ \bp } } \hat{h}^{\dagger}_{1,\bp }  \ket{ \AF }|^{2} +  |\bra{ \Psi^{\pm}_{ \bp } }  \hat{h}^{\dagger}_{2,\bp }  \ket{ \AF }|^{2} = Z^{\pm}_{\bp } .
\end{align}
Ultimately, this leaves the normalization condition to be 
\begin{align}
	a^{\pm}_{0}(\bp) =\sqrt{\frac{Z^{\pm}_{\bp }}{2}},
	\label{Eq:Norm}
\end{align}

\section{Reducing coefficients} \label{app:RedCoeff}

In this appendix, we will see that it is possible to define a single unique coefficient for each order using the recursion relation in Eq. \eqref{eq.recursive_relation_1}, the symmetries of the interaction vertices, Eq. \eqref{Eq:interaction_vertex_symmetry}, and those of the Green's functions
\begin{align}
	%\Sigma^{\mathrm{D}}(\mathbf{p}+\mathbf{Q},\omega) &= \Sigma^{\mathrm{D}}(\mathbf{p},\omega) \ , \ \Sigma^{\mathrm{O}}(\mathbf{p}+\mathbf{Q},\omega) = -\Sigma^{\mathrm{O}}(\mathbf{p},\omega)  \nonumber \\
	G_{\rm{d}}(\mathbf{p}+\mathbf{Q},\omega) &= G_{\rm{d}}(\mathbf{p},\omega) \nonumber\\
	 G_{\rm{o}}(\mathbf{p}+\mathbf{Q},\omega) &= -G_{\rm{o}}(\mathbf{p},\omega).
	\label{Eq:self_greens_symmetry}
\end{align}
The recursion relations can be restated as 
\begin{align}
	\begin{bmatrix}
   	a^{(1)}_{\{\mu_{n}\}}(\bp ,\{ \bk_{n+1} \}) \\
	a^{(2)}_{\{\mu_{n}\}}(\bp ,\{ \bk_{n+1} \})  
	\end{bmatrix} = \ \prod^{n-1}_{i=0}\mathcal{V}_{\{\mu_{n-i}\}}(\{ \bk_{n-i}\}; \bp, \epsilon^{\pm}_{\bp}) \cdot \begin{bmatrix}
   	a^{\pm}_{0}(\bp) \\
	\pm a^{\pm}_{0}(\bp)  
	\end{bmatrix}
\end{align}
with
\begin{align}
	&\mV_{\{ \mu_{n} \} }( \{ \bk_{n} \};\bp,\omega)= G({\bf K}_{n},\Omega^{\{ \bk_{n} \}}_{\{\mu_{n}\}}) V^{\dagger}_{\mu_{n}}({\bf K}_{n-1},\bk_{n}),
\end{align}
and $a^{(1)}(\bp) \equiv a^{\pm}_0(\bp)$ being the zeroth order coefficient.
To relate the different coefficients we effectively need to change all subscripts of $\mV$ to $\mu=+$. If translating the last excited spin waves crystal momentum by $\bQ$, we find 
\begin{multline}
	\mV_{\{\mu_{n-1}\},-}(\{ \bk_{n-1} \}, \bk_{n} + \bQ;\bp,\omega) = \\ 
	\begin{bmatrix}
   		 1 & 0  \\ 
		0  & -1
	\end{bmatrix} \cdot \mV_{\{\mu_{n-1}\},+}(\{ \bk_{n-1} \}, \bk_{n} ;\bp,\omega).
	\label{Eq:Coef_change_0}
\end{multline}
Hence, we see that by translating the momentum by $\bQ$ it is possible to change the subscripts at the expense of a possible sign. This relation also explains the signs in the expression for the wave function in Eq. \eqref{Eq:Wave_final}. If instead, we wish to change not the last subscript but any other, then we find 
 \begin{multline}
	\mV_{\{\mu_{n}\},-,\{\mu_{m}\}}(\{ \bk_{n} \}, \bk_{n+1} + \bQ,\{ \bk_{m} \};\bp,\omega) = \\
	\begin{bmatrix}
   		 1 & 0  \\ 
		0  & -1
	\end{bmatrix} \cdot \mV_{\{\mu_{n}\},+,\{\mu_{m}\}}(\{ \bk_{n} \}, \bk_{n+1},\{ \bk_{m} \};\bp,\omega) \cdot \begin{bmatrix}
   		 -1 & 0  \\ 
		0  & 1
	\end{bmatrix}.
	\label{Eq:Coef_change_1}
\end{multline}

If defining the coefficient with the hole present in layer 1 with excited spin waves only in the $+$ branch as
\begin{align}
	a^{(1)}_{+,+,...,+}( \bp , \{ \bk_{n} \} ) &\equiv a^{\pm}_{n}( \bp , \{ \bk_{n} \} ),
\label{eq.a_1_plusses_definition}
\end{align}
then it becomes possible with Eq. \eqref{Eq:Coef_change_1} and \eqref{Eq:Coef_change_0} to relate any $n$th order coefficient to $a^{\pm}_{n}$ by translating all momenta relating to a spin wave of the type $\mu=-$ by $\bk \rightarrow \bk + \bQ$. They will be related up to a sign which depends on the layer the hole is in, the types of excited spin waves, and the order in which they are excited. As an example we find $a^{(1)}_{+,-}(\bk_{1}, \bk_{2}+\bQ; \bp, \epsilon^{\pm}_{\bp}) = -a^{(1)}_{-,+}(\bk_{1}+\bQ, \bk_{2}; \bp, \epsilon^{\pm}_{\bp}) = -a^{\pm}_{2}(\bk_{1}, \bk_{2}; \bp, \epsilon^{\pm}_{\bp}) $. This also leaves it possible to write the recursion relation as the scalar relation 
\begin{align}
	a^{\pm}_{n+1}( \bp ,\{ \bk_{n+1} \} ) =& \ R^{\pm}(\bp ,\{ \bk_{n+1} \}) a^{\pm}_{n}(\bp, \{ \bk_{n} \} ),
	\label{Eq:RecRel}
\end{align} 
with the recursive coupling
\begin{align}
	R^{\pm}(\bp ,\{ \bk_{n} \}) =& \left( G_{\rm d}( \bK_{n},  \Omega^{ \pm }_{ \{ \bk_{n} \} })  \pm G_{\rm o}( \bK_{n} , \Omega^{ \pm }_{ \{ \bk_{n} \} } )  \right) \nonumber \\
	& \left( g_{+}\left( \bK_{n-1} ,\bk_{n}\right)  \pm f_{+}\left( \bk_{n}\right)  \right) \nonumber \\
	 \Omega^{ \pm }_{ \{ \bk_{n} \} }(\bp) =& \ \epsilon^{\pm}_\bp - \sum_{i=1}^{n}\omega_{+,\bk_{n}}.
	\label{Eq:Omega}
\end{align}

From the above relations, we will now derive relations between amplitudes in the wave function that will be necessary for the derivations of the B- and C-series. If changing the order of two different types of spin waves we get
\begin{multline}
	a^{(l)}_{\{\mu_{n}\},+,-,\{\mu_{m}\} }(\{\bk_{n}\},\bk,\bq+\bQ,\{\bk_{m}\}; \bp, \omega) =\\
	-a^{(l)}_{\{\mu_{n}\},-,+,\{\mu_{m}\} }(\{\bk_{n}\},\bk+\bQ,\bq,\{\bk_{m}\}; \bp, \omega).
	\label{Eq:Coef_change_1}
\end{multline}
If we instead wish to change only one subscript from $-$ to $+$, we find
\begin{multline}
	a^{(l)}_{\{\mu_{n}\},-,\{\mu_{m}\} }(\{\bk_{n}\},\bk+\bQ,\{\bk_{m}\}; \bp, \omega) =\\
	(-1)^{m+l}a^{(l)}_{\{\mu_{n}\},+,\{\mu_{m}\} }(\{\bk_{n}\},\bk,\{\bk_{m}\}; \bp, \omega).
	\label{Eq:Coef_change_2}
\end{multline}
At last, we investigate what happens if changing the first or two first types of spin waves. If changing just the first
\begin{multline}\label{Eq:Alt_Sign}
	a^{l}_{-,\{ \mu_{n-1} \}}(\bk_{1},\{ k_{n-1}\}; \bp, \omega)  =   \\
	(-1)^{n+l} a^{l}_{+,\{ \mu_{n-1} \}}(\bk_{1}+\bQ,\{ k_{n-1}\}; \bp, \omega). 
\end{multline}
For the first two

\begin{multline}
	a^{(l)}_{-,-,\{ \mu_{n-2} \}}(\bk_{1},\bk_{2},\{ k_{n-2}\}; \bp, \omega)  =  \\
	- a^{(l)}_{+,+,\{ \mu_{n-2} \}}(\bk_{1}+\bQ,\bk_{2}+\bQ, \{ k_{n-2}\}; \bp, \omega) \nonumber 
\end{multline}
\begin{multline}
	a^{(l)}_{+,-,\{ \mu_{n-2} \}}(\bk_{1},\bk_{2},\{ k_{n-2}\}; \bp, \omega) =  \\
	(-1)^{n+l} a^{(l)}_{+,+,\{ \mu_{n-2} \}}(\bk_{1},\bk_{2}+\bQ,\{ k_{n-2}\}; \bp, \omega) \nonumber 
\end{multline}
\begin{multline}\label{Eq:Alt_Sign_2}
	a^{(l)}_{-,+,\{ \mu_{n-2} \}}(\bk_{1},\bk_{2},\{ k_{n-2}\}; \bp, \omega) =  \\
	- (-1)^{n+l} a^{(l)}_{+,+,\{ \mu_{n-2} \}}(\bk_{1}+\bQ,\bk_{2}, \{ k_{n-2}\}; \bp, \omega).
\end{multline}

\section{B-series} \label{app:BSeries}

To understand the structure of this series, we will start by writing the wave function in a diagrammatic form. This form will not explicitly state the full complexity of the wave function and extra rules are needed to translate it into the different terms. This is though to work as a guide in understanding the structure 
\begin{align}
	\includegraphics{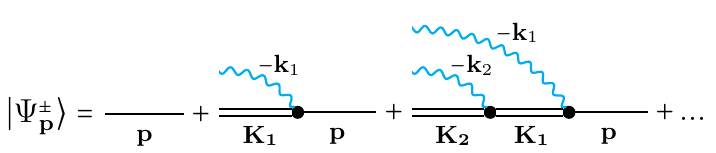}
\end{align}
The double line combined with the vertex translates to $R^{\pm}(\bp;\bk)$ while the straight line to $\sqrt{ Z^{\pm}_{\bp} }$. Looking at Eq. \eqref{Eq:Wave_final} we see that this is not the whole story since exciting $\mu = -$ spin wave leaves $R$ to be evaluated at $R(\bp,\bk+\bQ)$ and then we have a minus sign for specific vertices. 
We will come back to these features later, but for now, let us focus on the structure of the $B$ series. Looking at Eq. \eqref{Eq:B_series} we see that the diagrammatic structure of $B^{\pm,l}(\bq_{1},\bq_{2}; \bp, \omega )$ becomes
\begin{align}
	\includegraphics{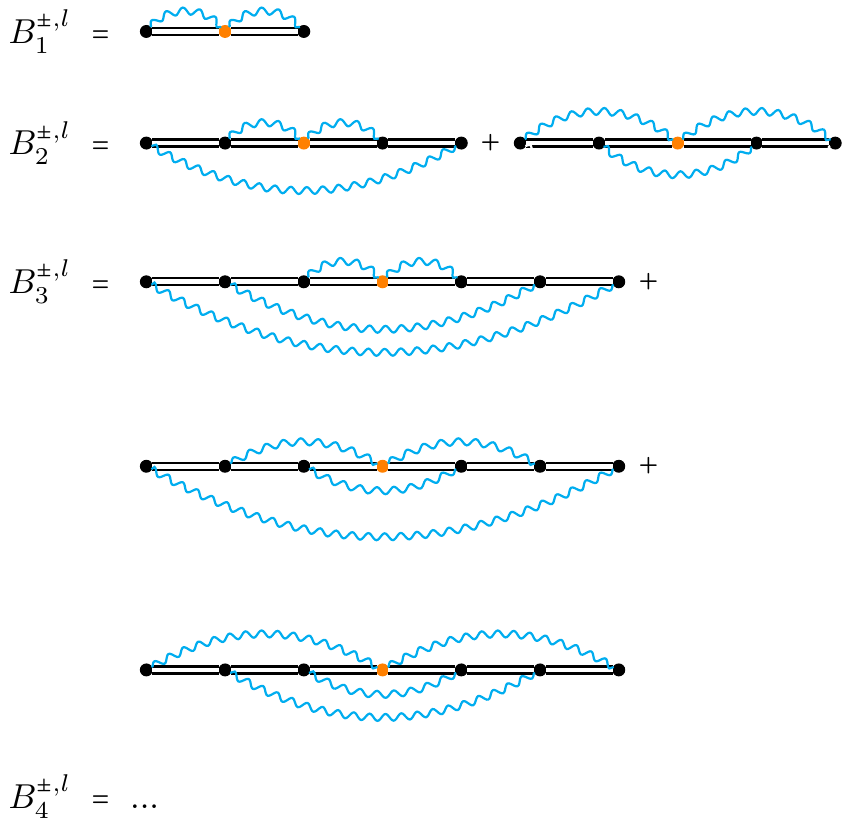}
\end{align}
where $B^{\pm,l} = Z^{\pm}_{\bp}\sum_{i=1}^{\infty}B^{\pm,l}_{i}$, and the orange vertex describes the $b_{\mu_{1},\mu_{2}}$ factor. The factor $Z^{\pm}_{\bp}$ comes from omitting the lines at the end in the diagrams. There will also be diagrams that are not symmetric around the center, $B^{asym}$, it is though seen in Appendix \ref{app:Vanish} that $B^{asym}=0$. To handle these diagrams we collect all the diagrams of the last type in $B^{\pm,l}_{0}$ such that
\begin{align}	
	\includegraphics{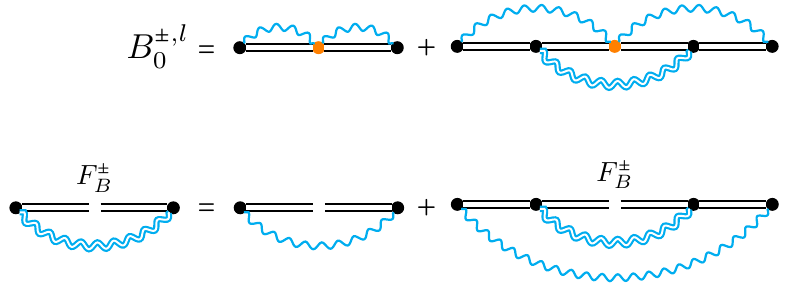}
	\label{Eq:B_0_diagram}
\end{align}
With these definitions, we see that it is possible to rewrite the $B$ series as
\begin{align}	
	\includegraphics{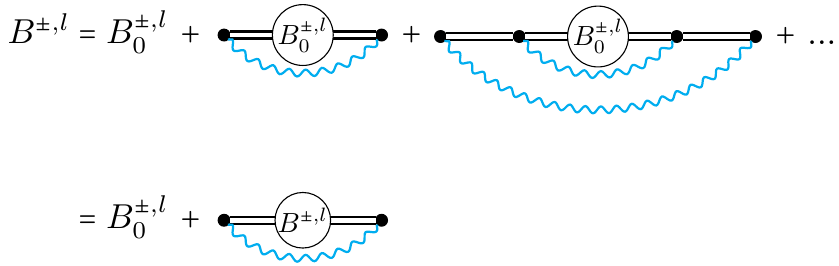}
	\label{Eq:B_Diagrammatic}
\end{align}
With these relations, we will start translating the diagrammatic representation. We will begin with $F^{\pm}_{B}$ where we have to remember, that the missing vertex in the center means that the momentum in the two propagators need not be the same. We also need to worry about the relation in Eq. \eqref{Eq:Alt_Sign}. This says, that it actually matters whether or not the diagrams attached to $F_{B}$ contain two spin-wave of the same kind or opposite. If they are opposite then the diagrams need to come with alternating signs according to Eq. \eqref{Eq:Alt_Sign}. For now, we will disregard the factor of $(-1)^{l}$ to find the following for the same kinds of spin waves
\begin{align}
	F^{\pm,s}_{B}(\bq_{1},\bq_{2};\bp,\omega)  =   \sum_{\bk}  \big(  R^{\pm}(\bq_{1},\bk; \bp,\omega) R^{\pm}(\bq_{2},\bk; \bp, \omega) \times \nonumber  \\   
	\left[ 1 + F^{\pm,s}_{B}(\bq_{1},\bq_{2};\bp+\bk,\omega-\omega^{+}_{\bk})  \right] + 
	 R^{\pm}(\bq_{1},\bk+\bQ; \bp, \omega) \times \nonumber  \\ 
	R^{\pm}(\bq_{2},\bk+\bQ; \bp, \omega)  \left[ 1 + F^{\pm,s}_{B}(\bq_{1},\bq_{2};\bp+\bk ,\omega-\omega^{-}_{\bk})  \right] \big),
	\label{Eq:F_B_1_s}
\end{align}
and for two different 
\begin{align}
	F^{\pm,a}_{B}(\bq_{1},\bq_{2};\bp,\omega)  =  - \sum_{\bk}  \big(  R^{\pm}(\bq_{1},\bk; \bp,\omega) \times \nonumber  \\ 
	R^{\pm}(\bq_{2},\bk; \bp, \omega) \left[ 1 + F^{\pm,a}_{B}(\bq_{1},\bq_{2};\bp+\bk,\omega-\omega^{+}_{\bk})  \right] + \nonumber  \\ 
	  R^{\pm}(\bq_{1},\bk+\bQ; \bp, \omega) R^{\pm}(\bq_{2},\bk+\bQ; \bp, \omega) \times \nonumber  \\ 
	\left[ 1 + F^{\pm,a}_{B}(\bq_{1},\bq_{2};\bp+\bk ,\omega-\omega^{-}_{\bk})  \right] \big),
	\label{Eq:F_B_1_a}
\end{align}
with the recursive coupling now being defined as
\begin{align}
	R^{\pm}(\{ \bk_{n} \};\bp,\omega) =& \left( G_{\rm d}( \bK_{n},  \Omega^{ \pm }_{ \{ \bk_{n} \} })  \pm G_{\rm o}( \bK_{n} , \Omega^{ \pm }_{ \{ \bk_{n} \} } )  \right) \nonumber \\
	& \left( g_{+}\left( \bK_{n-1} ,\bk_{n}\right)  \pm f_{+}\left( \bk_{n}\right)  \right) \nonumber \\
	 \Omega^{ \pm }_{ \{ \bk_{n} \} }(\omega) =& \ \omega - \sum_{i=1}^{n}\omega_{+,\bk_{n}},
	\label{Eq:RecursiveCoupling}
\end{align}
in order to encompass being evaluated away from the quasi-particle energies. As a sanity check, we find that in the limit of $t_{\perp} = 0$ the results are equivalent to those obtained for a single layer \cite{nielsen2021}. \\

With this, we can now approach $B^{\pm,l}_{0}$, where we need to remember that the two boson lines in the first term in Eq. \eqref{Eq:B_0_diagram} need not be the same and the vertex $b_{\mu_{1},\mu_{2}}(\bq_{1},\bq_{2})$ depends on the types of spin waves. The fact that this diagram also allows for two different kinds of spin waves makes it important to think about which layer the hole is present in. Looking at Eq. \eqref{Eq:Wave}, we see that if the hole is present in layer 1, then the two types of spin waves come with the same sign while for layer 2 this is not the case. This means that $B^{\pm,l}_{1}$ comes with a $-(-1)^{l}$ in front when the two spin waves are of the opposite kind. Having this in mind we find 
\begin{align}
	B^{\pm,l}_{0}(\bq_{1},\bq_{2}; \bp, \omega) = \frac{Z^{\pm}_{\bp}}{2}\big[ R^{\pm}(\bq_{1}; \bp, \omega) R^{\pm}(\bq_{2}; \bp, \omega) \times \nonumber \\  
	b_{+,+}(\bq_{1},\bq_{2}) \cdot  \big[ 1 + F^{\pm,s}_{B}(\bq_{1},\bq_{2};\bp,\omega)  \big]   +   \nonumber \\ 
	R^{\pm}(\bq_{1}+\bQ; \bp, \omega) R^{\pm}(\bq_{2}+\bQ; \bp, \omega)b_{-,-}(\bq_{1},\bq_{2})  \times \nonumber \\  
	\big[ 1 + F^{\pm,s}_{B}(\bq_{1}+\bQ,\bq_{2}+\bQ;\bp,\omega)  \big] -(-1)^{l}R^{\pm}(\bq_{1}; \bp, \omega)\times \nonumber \\ 
	R^{\pm}(\bq_{2}+\bQ; \bp, \omega)b_{+,-}(\bq_{1},\bq_{2}) \cdot  \big[ 1 + F^{\pm,a}_{B}(\bq_{1},\bq_{2}+\bQ;\bp,\omega)  \big]  \nonumber \\ 
	  - (-1)^{l}R^{\pm}(\bq_{1}+\bQ; \bp, \omega)R^{\pm}(\bq_{2}; \bp, \omega)b_{-,+}(\bq_{1},\bq_{2})  \big] \times \nonumber \\   
	  \big[ 1 + F^{\pm,a}_{B}(\bq_{1}+\bQ,\bq_{2};\bp,\omega)  \big]   
	\label{Eq:B_0}
\end{align}
Here, we find $B^{\pm,2}=0$ in the limit of $t_{\perp}/t = 0$ and $B^{\pm,1}$ to coincide with the result for a single layer. This is expected since in Eq. \eqref{Eq:Mag_l_2} we choose the spin to be measured in layer 1, it will therefore make sense that if choosing $l=1$ and taking the limit of $t_{\perp}=0$ we expect to find $M^{1,2}=0$, $B^{\pm,1}=0$ and $C^{\pm,1}=0$. Another feature found in Eq. \eqref{Eq:B_0} is that the only difference between the two layers comes from the terms mixing the spin wave branches. If we take the extreme limit of $t_{\perp}\gg t$ we have $b_{+,-}=0=b_{-,+}$. The same holds for the $C$-series, such that all states, as expected, will show perfect antialignment between the layers in this limit. It is not trivial what happens for intermediate values [see Fig. \ref{Fig:DeltaMag}].

With these relations, we can now present the translation of Eq. \eqref{Eq:B_Diagrammatic}
\begin{align}
	B^{\pm,l}(\bq_{1},\bq_{2}; \bp, \omega) = B^{\pm,l}_{0}(\bq_{1},\bq_{2}; \bp, \omega) + \nonumber \\
	\sum_{\bk} \big[ R^{\pm}(\bk; \bp, \omega) R^{\pm}(\bk; \bp, \omega) \cdot B^{\pm,l}(\bq_{1},\bq_{2}; \bp + \bk, \omega - \omega^{+}_{\bk}) + \nonumber \\
	 R^{\pm}(\bk+\bQ; \bp, \omega) R^{\pm}(\bk+\bQ; \bp, \omega)   \cdot B^{\pm,l}(\bq_{1},\bq_{2}; \bp+\bk, \omega - \omega^{-}_{\bk}) \big] 
	\label{Eq:B}
\end{align}
We can now find $B^{\pm,l}$ by first solving Eq. \eqref{Eq:F_B_1_s} and \eqref{Eq:F_B_1_a} self-consiestently starting from $F^{\pm}_{B}=0$. The results are then inserted into Eq. \eqref{Eq:B_0} which is then used to solve Eq. \eqref{Eq:B} self-consistently starting from  $B^{\pm,l} = B^{\pm,l}_{0}$.

\section{C-series} \label{app:CSeries}

Finding an expression for the $C$ series is much like finding the one for the $B$ series. The $C$ series differs in that it will connect different orders of the wave function while the $B$ series connects the same orders. Again, we define $C = \sum_{n=0}^{\infty}C_{n,n+2}$, where $C_{i,j}$ represents the overlap between the $i^{th}$ order in $\bra{\Psi^{\pm}_{\bp}}$ and the $j^{th}$ order in $\ket{\Psi^{\pm}_{\bp}}$. $C_{n,n+2}$ can diagrammatically be represented as
\begin{align}
	\includegraphics{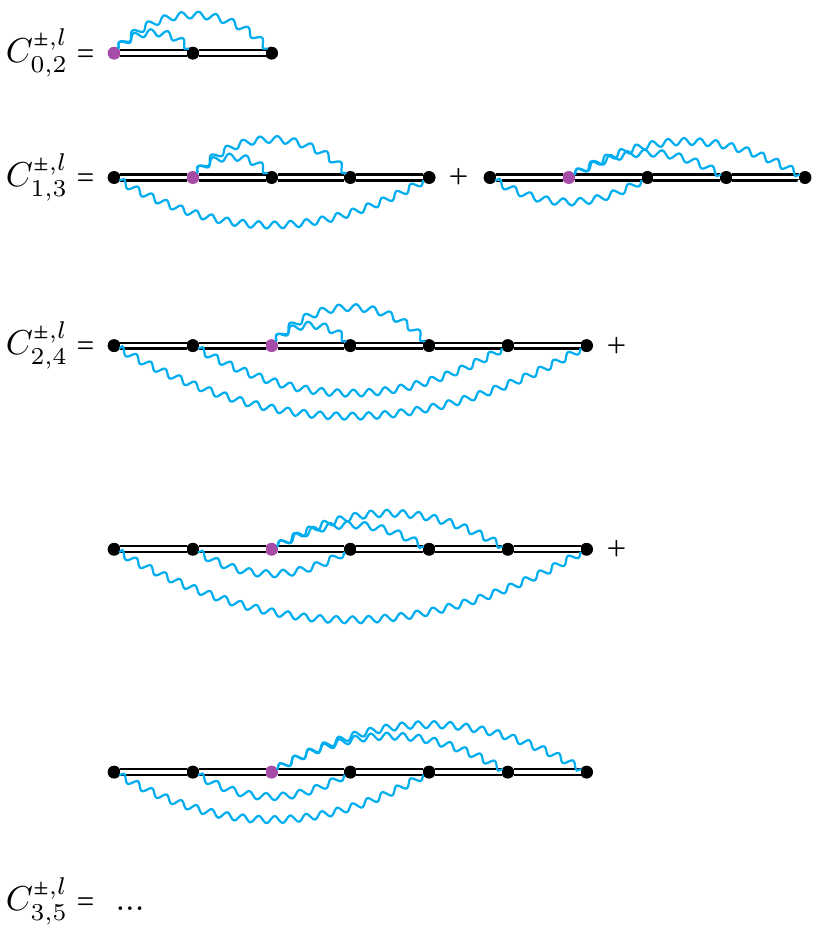}
\end{align}
We also have other diagrams, but we find them summing to zero (See Appendix \ref{app:Vanish}). If collecting the last type of diagrams in the quantity $C_{0}$, we see
\begin{align}
	\includegraphics{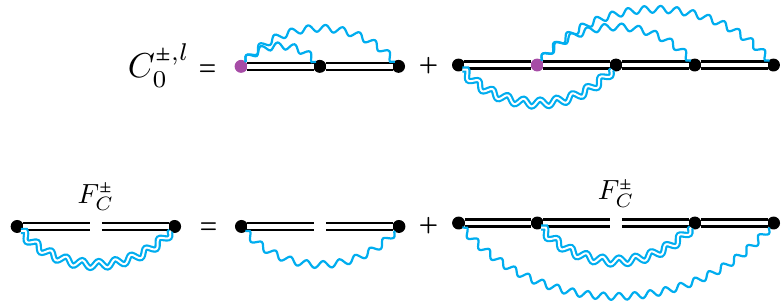}
	\label{Eq:C_0}
\end{align}
The derivation of $F_{C}$ follows that of $F_{B}$ with the only difference being, that in the expression for $C^{\pm,l}_{0} $ we have two spin waves excited to the right of $F_{C}$ and none to the left, for $F_{B}$ we had one on each side. We, therefore, need to use the relations presented in Eq. \eqref{Eq:Alt_Sign_2}. Again we need to define two types of $F_{C}$ where one has alternating signs
\begin{align}
	F^{\pm,s}_{C}(\bq_{1},\bq_{2};\bp,\omega) = \sum_{\bk} \big[  R^{\pm}(\bk; \bp, \omega) R^{\pm}(\bq_{1},\bq_{2},\bk; \bp, \omega) \times \nonumber \\ 
	\left[ 1 + F^{\pm,s}_{C}(\bq_{1},\bq_{2};\bp+\bk,\omega-\omega^{+}_{\bk})  \right]  +  \nonumber \\ 
	R^{\pm}(\bk+\bQ; \bp, \omega)R^{\pm}(\bq_{1},\bq_{2},\bk+\bQ; \bp, \omega) \big[ 1 + \nonumber \\
	F^{\pm,s}_{C}(\bq_{1},\bq_{2};\bp+\bk,\omega-\omega^{-}_{\bk})  \big] 
	\nonumber
\end{align}
\begin{align}
	F^{\pm,a}_{C}(\bq_{1},\bq_{2};\bp,\omega) = -\sum_{\bk} \big[  R^{\pm}(\bk; \bp, \omega) R^{\pm}(\bq_{1},\bq_{2},\bk; \bp, \omega) \times \nonumber \\ 
	\left[ 1 + F^{\pm,a}_{C}(\bq_{1},\bq_{2};\bp+\bk,\omega-\omega^{+}_{\bk})  \right]  +  \nonumber \\ 
	R^{\pm}(\bk+\bQ; \bp, \omega)R^{\pm}(\bq_{1},\bq_{2},\bk+\bQ; \bp, \omega) \big[ 1 + \nonumber \\
	F^{\pm,a}_{C}(\bq_{1},\bq_{2};\bp+\bk,\omega-\omega^{-}_{\bk})  \big].
	\label{Eq:F_Ca}
\end{align}
Here, we again use the definition for the recursive coupling given in Eq. \eqref{Eq:RecursiveCoupling}. To find $C^{\pm,l}_{0}$ we have to be a bit careful since the purple vertex is seen as only acting on $\ket{\Psi^{\pm}_{\bp}}$. In the correlator, Eq. \eqref{Eq:C_0}, we have $\hat{b}_{\mu_{1},-\bq_{1} }\hat{b}_{\mu_{2},-\bq_{2} }$ such that $\hat{b}_{\mu_{1},-\bq_{1} }$ can either annihilate the first or the second spin wave. Using that $c_{\mu_{1},\mu_{2}}(\bq_{1},\bq_{2})=c_{\mu_{2},\mu_{1}}(\bq_{2},\bq_{1})$ this will yield a factor of two and we thereby find
\begin{align}
	C^{\pm,l}_{0}(\bq_{1},\bq_{2}; \bp, \omega) =  \ Z^{\pm}_{\bp} \big( R^{\pm}(\bq_{1}; \bp, \omega) R^{\pm}(\bq_{1},\bq_{2}; \bp, \omega) \times \nonumber \\ 
	c_{+,+}(\bq_{1},\bq_{2})  \cdot \big[ 1 + F^{\pm,s}_{C}(\bq_{1},\bq_{2};\bp,\omega)  \big] \nonumber \\ 
	- R^{\pm}(\bq_{1}+\bQ; \bp, \omega) R^{\pm}(\bq_{1}+\bQ,\bq_{2}+\bQ; \bp, \omega)c_{-,-}(\bq_{1},\bq_{2})  \times \nonumber \\  \big[ 1 + F^{\pm,s}_{C}(\bq_{1}+\bQ,\bq_{2}+\bQ;\bp,\omega)  \big] \nonumber \\ 
	-(-1)^{l} R^{\pm}(\bq_{1}; \bp, \omega) R^{\pm}(\bq_{1},\bq_{2}+\bQ; \bp, \omega)c_{+,-}(\bq_{1},\bq_{2})  \times \nonumber \\  \big[ 1 + F^{\pm,a}_{C}(\bq_{1},\bq_{2}+\bQ;\bp,\omega)  \big] \nonumber \\ 
	+ (-1)^{l} R^{\pm}(\bq_{1}+\bQ; \bp, \omega) R^{\pm}(\bq_{1}+\bQ,\bq_{2}; \bp, \omega)c_{-,+}(\bq_{1},\bq_{2}) \big] \times \nonumber \\ \big[ 1 + F^{\pm,a}_{C}(\bq_{1}+\bQ,\bq_{2};\bp,\omega)  \big]\big). 
\end{align}
Here, the factors in front are a bit ambiguous, but they can trivially be derived from in Eq. \eqref{Eq:Alt_Sign_2}. As for the $B$-series we find $c_{+,-}=0=c_{-,+}$ in the extreme limit of $t_{\perp}\gg t$, thereby, leaving the spins perfectly antialigned between the layers.

With the definition of $C^{\pm,l}_{0} $ we can write the $C$ series as
\begin{align}
	\includegraphics{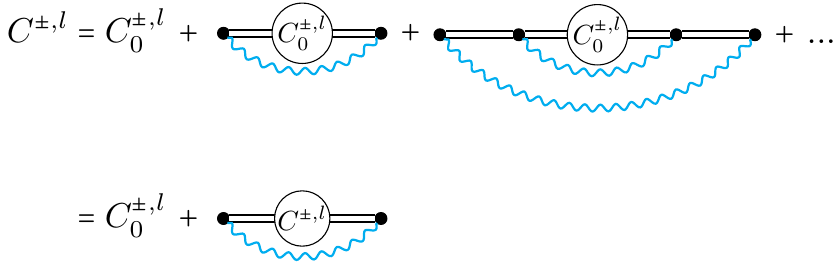}
	\label{Eq:C_Diagrammatic}
\end{align}
This translates to 
\begin{multline}
	C^{\pm,l}(\bq_{1},\bq_{2}; \bp, \omega) =  \ C^{\pm,l}_{0}(\bq_{1},\bq_{2}; \bp, \omega) +  \nonumber \\
	 \sum_{\bk} \big[ R^{\pm}(\bk; \bp, \omega) R^{\pm}(\bk; \bp, \omega) \cdot C^{\pm,l}(\bq_{1},\bq_{2}; \bp + \bk, \omega - \omega^{+}_{\bk}) + \nonumber \\
	 R^{\pm}(\bk+\bQ; \bp, \omega) R^{\pm}(\bk+\bQ; \bp, \omega)   \cdot C^{\pm,l}(\bq_{1},\bq_{2}; \bp+\bk, \omega - \omega^{-}_{\bk}) \big] 
\end{multline}
As a sanity check, we find that for $t_{\perp}\rightarrow0$ this yields the result presented in \cite{nielsen2021} for a single layer. 

\section{Vanishing diagrams} \label{app:Vanish}

In this section, we will show that the nonsymmetric diagrams for the $B$ series and the rest of the diagrams for the $C$ series vanish. The structure of the nonvanishing diagrams is such that for the $B$ series they will not be symmetric around the center. For the $C$ series, the two spin waves annihilated by the purple vertex will not be excited one after the other. An example from both series is stated here
\begin{align}
	\includegraphics{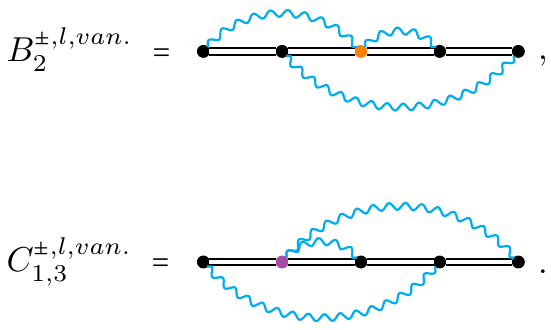}
	\label{Eq:Ex_vanish}
\end{align}
The feature determining whether the diagrams vanish or not is which terms in the diagrams come with the same vertex factor, so $b_{\mu_{1},\mu_{2}}$ or $c_{\mu_{1},\mu_{2}}$. This is not obvious, we, therefore, examine the diagrams from Eq. \eqref{Eq:Ex_vanish}. 

For $B^{\pm,l,van.}_{2} $ the diagram translates to
\begin{widetext}
	\begin{align}
	B^{\pm,l,van.}_{2} & (\bq_{1},\bq_{2}; \bp, \omega) = \ \frac{Z^{\pm}_{\bp}}{2}\big[ \nonumber \\
	& +b_{+,+}(\bq_{1},\bq_{2}) R^{\pm}(\bq_{1}; \bp, \omega) \sum_{\bk} R^{\pm}(\bq_{1},\bk; \bp, \omega)R^{\pm}(\bk,\bq_{2}; \bp, \omega)R^{\pm}(\bk; \bp, \omega)\nonumber \\ 
	& - b_{+,+}(\bq_{1},\bq_{2})  R^{\pm}(\bq_{1}; \bp, \omega) \sum_{\bk} R^{\pm}(\bq_{1},\bk+\bQ; \bp, \omega)R^{\pm}(\bk+\bQ,\bq_{2}; \bp, \omega)R^{\pm}(\bk+\bQ; \bp, \omega)\nonumber \\ 
	& -(-1)^{l} b_{+,-}(\bq_{1},\bq_{2}) R^{\pm}(\bq_{1}; \bp, \omega) \sum_{\bk} R^{\pm}(\bq_{1},\bk; \bp, \omega)R^{\pm}(\bk,\bq_{2}+\bQ; \bp, \omega)R^{\pm}(\bk; \bp, \omega)\nonumber \\ 
	& +(-1)^{l} b_{+,-}(\bq_{1},\bq_{2}) R^{\pm}(\bq_{1}; \bp, \omega) \sum_{\bk} R^{\pm}(\bq_{1},\bk+\bQ; \bp, \omega)R^{\pm}(\bk+\bQ,\bq_{2}+\bQ; \bp, \omega)R^{\pm}(\bk+\bQ; \bp, \omega) \nonumber \\ 
	& +(-1)^{l} b_{-,+}(\bq_{1},\bq_{2}) R^{\pm}(\bq_{1}+\bQ; \bp, \omega) \sum_{\bk} R^{\pm}(\bq_{1}+\bQ,\bk; \bp, \omega)R^{\pm}(\bk,\bq_{2}; \bp, \omega)R^{\pm}(\bk; \bp, \omega) \nonumber \\ 
	& -(-1)^{l} b_{-,+}(\bq_{1},\bq_{2}) R^{\pm}(\bq_{1}+\bQ; \bp, \omega) \sum_{\bk} R^{\pm}(\bq_{1}+\bQ,\bk+\bQ; \bp, \omega)R^{\pm}(\bk+\bQ,\bq_{2}; \bp, \omega)R^{\pm}(\bk+\bQ; \bp, \omega) \nonumber \\ 
	& - b_{-,-}(\bq_{1},\bq_{2}) R^{\pm}(\bq_{1}+\bQ; \bp, \omega) \sum_{\bk} R^{\pm}(\bq_{1}+\bQ,\bk; \bp, \omega)R^{\pm}(\bk, \bq_{2}+\bQ; \bp, \omega)R^{\pm}(\bk; \bp, \omega)\nonumber \\ 
	& + b_{-,-}(\bq_{1},\bq_{2}) R^{\pm}(\bq_{1}+\bQ; \bp, \omega) \sum_{\bk} R^{\pm}(\bq_{1}+\bQ,\bk+\bQ; \bp, \omega)R^{\pm}(\bk+\bQ, \bq_{2}+\bQ; \bp, \omega)R^{\pm}(\bk+\bQ; \bp, \omega) \big]
	\label{Eq:B_van}
	\end{align}
\end{widetext}
Here we see that by taking $\bk \rightarrow \bk + \bQ$ for every second term this diagram yields zero. The recurring feature leading to these terms being zero, in comparison to the symmetric version, is that the terms coming with the same vertex amplitude, $b_{\mu_{1},\mu_{2}}$, can be put into pairs of the kind $a_{\{\mu_{n}\}, \mu_{1}, \mu,\{\mu_{m}\} }\cdot a_{\{\mu_{n}\}, \mu, \mu_{2},\{\mu_{m}\} }$ and $a_{\{\mu_{n}\}, \mu_{1}, \bar{\mu},\{\mu_{m}\} }\cdot a_{\{\mu_{n}\}, \bar{\mu}, \mu_{2},\{\mu_{m}\} }$, where $\bar{\mu}$ is the opposite type of spin wave to $\mu$, and the associated momentum is summed over. If looking at Eq. \eqref{Eq:Coef_change_1}, we see that such terms come with an opposite sign. Now, since the moment associated with $\mu_{2}$ is summed over, we can translate them by $\bQ$ such that the two terms will sum up to zero. The equivalent symmetric diagrams will instead come with the same sign in front.

For the C series, the vanishing diagrams yield zero for similar reasons. Here, terms coming with the same vertex factor can be grouped in pairs of $a_{\{ \mu_{n} \},\mu,\{ \mu_{m} \}} \cdot a_{\{ \mu_{n},\mu_{1} \},\mu,\mu_{2},\{ \mu_{m} \}} $ and $a_{\{ \mu_{n} \},\bar{\mu},\{ \mu_{m} \}} \cdot a_{\{ \mu_{n},\mu_{1} \},\bar{\mu},\mu_{2},\{ \mu_{m} \}} $, where the momentum associated with $\mu$ is summed over. One should understand $\{ \mu_{n},\mu_{2} \}$ as series of excited spin waves where $\mu_{2}$ excited at some unspecified order. Looking at Eq. \eqref{Eq:Coef_change_2}, we see that these two terms come with a sign difference. We can therefore translate the momentum by $\bQ$ to have them cancel one another.

\section{Symmetries of the polaron cloud} \label{app:sym}
In this appendix we use the expression for the $B$- and $C$-series, Eq. \eqref{Eq:B_series} and \eqref{Eq:C_series}, to show the symmetry 
\begin{align}
 	M_{l}^{-}(\bd,\bp + \bQ) &= M_{l}^{+}(\bd,\bp).
 \end{align}
Looking at the recursive coupling, Eq. \eqref{Eq:Omega}, we find the following symmetry related to the AF ordering vector
\begin{align}
	R^{+}(\bp + \bQ,\{\bk_{n}\}) = -R^{-}(\bp,\{\bk_{n}\}).
\end{align}
In the expression for the $B$- and $C$-series, Appendix \ref{app:BSeries} and \ref{app:CSeries}, we find that all $R^{\pm}$ are multiplied by even numbers of $R^{\pm}$s. The same applies to the residue $Z^{\pm}_{\bp}$, such that this minus will change nothing and we retrieve $M_{l}^{+}(\bd,\bp) = M_{l}^{-}(\bd,\bp + \bQ)$. Realizing that $\Psi^{+}_{\bp + \bQ}$ and $\Psi^{-}_{\bp}$ are degenerate states, see Ref.~\cite{nyhegn2022}, we find that this manifests itself in the same frustration of the magnetic environment. 

\section{Polaron cloud for $J_{\perp} \neq 4t^{2}_{\perp}/U$} \label{App:Decoupled}
\begin{figure} 
	\begin{center}
	\hspace{-1.5cm}
	\includegraphics[width=0.4\textwidth]{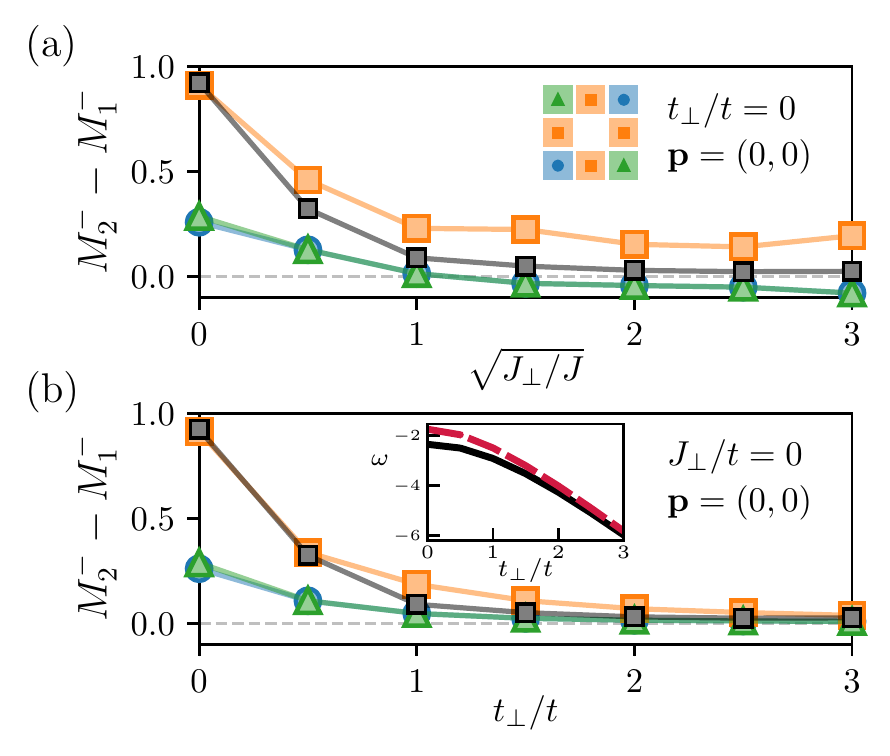}
	\end{center}
	\caption{Difference in magnetization between the layers for $J/t=0.3$ and $\bp =(0,0)$. (a) Here shown as a function of the interlayer spin-spin correlation with $t_{\perp}/t = 0$. $\sqrt{J_{\perp}/J}$ is chosen since $\sqrt{J_{\perp}/J} = t_{\perp}/t$ for $J_{\perp} = 4t^{2}_{\perp}/U$, and, therefore, enables easy comparison. In (b), we instead show it as a function of the interlayer hopping $t_{\perp}/t$ with $J_{\perp}/J=0$. The grey curves in both panels correspond to the nearest neighbor difference when $J_{\perp} = 4t^{2}_{\perp}/U$, and should be compared to the orange curve. Doing so, we see that in (a) the mirroring effect is less prominent while in (b) it follows more closely. The inset shows the energy of the $\bp=(\pi/2,\pi/2)$ state in black and $\bp=(0,0)$ in red for $J_{\perp}/J=0$. The $\bp=(\pi/2,\pi/2)$ state, thus, remains the ground state with increasing interlayer hopping for $t_\perp = 0$. }
	\label{Fig:DeltaMag_const}
\end{figure}

In this Appendix, we investigate the influence of $J_{\perp}$ and $t_{\perp}$ separately in relation to how the polaron clouds in the two layers mimic one another for large $J_{\perp}/t$. To do so, we deviate from $J_{\perp} = 4t^{2}_{\perp}/U$ used in the main text, such that the model no longer describes the low energy physics of the Fermi-Hubbard model in the large $U$ limit. 
In analog to Fig. \ref{Fig:DeltaMag}(a) in the main text, Fig. \ref{Fig:DeltaMag_const} show $M^{-}_1 - M^{-}_2$ for $t_{\perp}/t=0$ (a) and $J_{\perp}/t=0$ (b) with $\bp = (0,0)$. The grey curves show the nearest neighbor results plotted in the main text for $J_{\perp} = 4t^{2}_{\perp}/U$. Comparing the orange and grey curves in Fig. \ref{Fig:DeltaMag}(a), we see that a vanishing interlayer hopping, $t_\perp / t = 0$, yields a less prominent mirroring effect between the layers. In Fig. \ref{Fig:DeltaMag}(b), for $J_{\perp}/t=0$, the orange and grey curves follow each other more closely, though $M^{-}_1 - M^{-}_2$ stays larger with no interlayer spin-spin interaction.
The mirroring effect where the cloud upholds the interlayer AF order is, therefore, primarily created by interlayer hopping. That said, the energy decrease associated with keeping the AF order between the layers is absent in the case of $J_{\perp}/t\neq0$. As seen in the inset in Fig. \ref{Fig:DeltaMag}(b), the system will, consequently, not prefer $\bp=(0,0)$ as its ground state. As a result, in order for the system to prefer the $\bp=(0,0)$ polaron state as its ground state, we need the interlayer spin-spin interaction.

\section{Deviating from $J/t = 0.3$} \label{App:J_t}
\begin{figure}
	\begin{center}
	\includegraphics[width=0.4\textwidth]{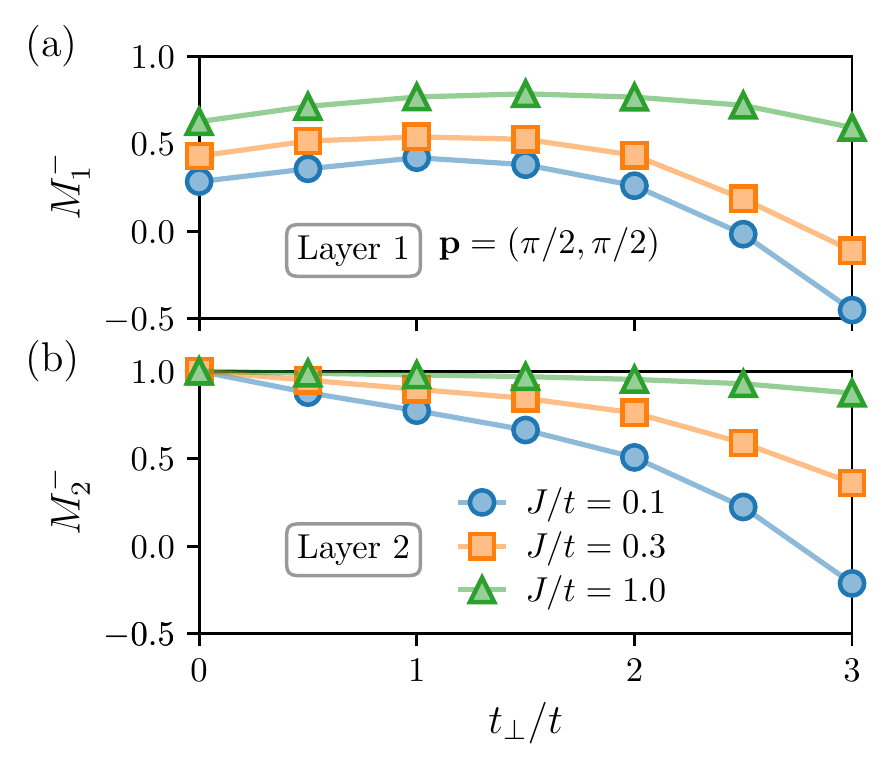}
	\end{center}
	\caption{This figure shows the frustration of the nearest neighboring site in layer 1 where the hole is present and the opposite sites in layer 2. This is plotted for $J/t=0.1,0.3,1.0$ with varying interlayer hopping, $t_{\perp}/t$. Here, we see the same behavior as described in the main text with the overall frustration being larger for smaller $J/t$.}
	\label{Fig:Mag_J}
\end{figure}

This appendix discusses the influence of the value $J/t$ on the magnetic dressing cloud. Similar to Fig. \ref{Fig:MagFrus_t_perp} in the main text, we show in Fig. \ref{Fig:Mag_J} the development of the nearest neighbor and next-nearest neighbor frustrations in layer 1 and 2, respectively, as a function of interlayer hopping for $J/t= 0.1,0.3,1.0$. Here, we see that they all follow the same behavior but with increasing frustration for smaller $J/t$. This is because for smaller $J/t$ the energy penalty of being localized increases and the hole will be more delocalized, thereby, increasing the frustration. The features described in the main text are independent of this battle between the two terms since it reflects the behavior of the order parameter dictated by $J_{\perp}/J$. By changing $J/t$ the behavior described in the main text remains qualitative the same, but will quantitatively take place with more ($J/t<0.3$) or less ($J/t>0.3$) overall frustration.
\clearpage

\bibliography{ArticleTheWave,ArticleTheWave2}

\end{document}